\documentclass[12pt,preprint]{aastex}

\shorttitle{Star Formation in the N. Complex of NGC~2264}
\shortauthors{Hedden et al.}

\begin{document}

\title{Star Formation in the Northern Cloud Complex of NGC~2264}
\author{Abigail S. Hedden, Christopher K. Walker, and Christopher E. Groppi}
\affil{Steward Observatory, University of Arizona, Tucson, AZ 85721}
\email{ahedden@as.arizona.edu}
\author{Harold M. Butner}
\affil{Joint Astronomy Centre, Hilo, HI 96720}

\begin{abstract}
We have made continuum and spectral line observations of several outflow sources in the Mon~OB1 dark cloud (NGC~2264) using the Heinrich Hertz Telescope (HHT) and ARO 12m millimeter-wave telescope.  This study explores the kinematics and outflow energetics of the young stellar systems observed and assesses the impact star formation is having on the surrounding cloud environment.  Our data set incorporates $^{12}$CO\,$(3-2)$, $^{13}$CO\,$(3-2)$, and $^{12}$CO\,$(1-0)$ observations of outflows associated with the sources IRAS~06382+1017 and IRAS~06381+1039, known as IRAS~25 and 27, respectively, in the northern cloud complex.  Complementary $870\, \micron$ continuum maps were made with the HHT 19 channel bolometer array.  Our results indicate there is a weak $\le\,0.5$\,\% coupling between outflow kinetic energy and turbulent energy of the cloud.  An analysis of the energy balance in the IRAS~25 and 27 cores suggests they are maintaining their dynamical integrity except where outflowing material directly interacts with the core, such as along the outflow axes.   
\end{abstract}

\keywords{stars: formation, ISM: individual - Mon~OB1 (NGC~2264) - Clouds - ISM: jets and outflows - ISM: kinematics and dynamics - ISM: molecules - radio lines: ISM}

\section{Introduction}

The Mon~OB1 dark cloud lies toward the galactic anticenter and has been the subject of several large-scale molecular line surveys \citep{bli79}, \citep{omt96}.  Due to its location, there is little source confusion from intervening foreground and background clouds.  The relative proximity of this complex, 800\,pc according to the work of \cite{walker56}, has also contributed to its survey appeal.  The region of Mon~OB1 associated with NGC~2264 has itself been the subject of several studies, including an unbiased $^{12}$CO\,$(1-0)$ survey and the discovery of nine molecular outflows \citep{ml85}, \citep{mls88}.  \cite{mly89} presented the results of an IRAS survey of 30 discrete far-infrared sources, including IRAS~25 and IRAS~27.  The IRAS Point Source Catalog (PSC) identifies IRAS~25 and IRAS~27 with objects IRAS~06382+1017 and IRAS~06381+1039, respectively.  A systematic search for dense gas with CS observations revealed that these objects were associated with a $\sim\,500-700$ M$_{\odot}$ clump \citep{wwl95}. 
              
\cite{wc03} used HIRES processing of IRAS data, $^{12}$CO\,$(2-1)$ observations, and SCUBA continuum images to identify individual cores embedded within NGC~2264 and concluded that IRAS~25 and IRAS~27, initially thought to be Class~I protostellar objects \citep{mly89}, consist of multiple sources in different evolutionary stages.  They also find evidence for multiple outflow driving sources in the $^{12}$CO$(2-1)$ data.  In a wide field imaging survey of the NGC~2264 region of Mon~OB1 in the infrared, H$_{\alpha}$, and $^{12}$CO\,$(3-2)$, \cite{rei04} report several new Herbig-Haro (HH) objects near the star forming cores found by \cite{wc03} and attempt to pinpoint their origins.  

A primary goal of this work is to determine physical characteristics and outflow properties of IRAS~25 and 27.  With this information, we hope to gain a better understanding of the star formation history, dynamics, and impact that these protostellar objects have on the surrounding cloud environment.  In Section (2.), we describe the observations presented in this work.  Section (3.) addresses the results of the continuum and spectral line observations of each of the sources, IRAS~25 and 27.  In Section (4.), we discuss velocity centroid studies of the molecular line data and draw energetics comparisons to assess the impact star formation is having upon the cloud complex.  A summary of our findings is presented in (5.)       

\section{Observations}

\subsection{Continuum mapping with the HHT}

The continuum data presented in this work were obtained with the facility 19 channel array operating at the HHT on Mt. Graham, Arizona.  This array, developed by E. Kreysa and collaborators, contains 19 individual broadband bolometric detectors sensitive around a central wavelength of $\lambda=870\, \micron$ (345\,GHz) and arranged in two concentric hexagonal rings surrounding a central pixel.  The field of view of the array is $\sim200''$ in diameter.  Observations were made during 2003 January $12-14^{th}$ and May $6-8^{th}$, and 2004 April $21-22^{nd}$ using On The Fly (OTF) mapping.  With this technique, fully sampled $450'' \times 300''$ maps were made surrounding the sources IRAS~25 and 27 in NGC~2264.  A total of six maps of IRAS~27 and three maps of IRAS~25 were combined to produce the data sets presented here.  During observations, the pointing was checked each hour by observing a planet, such as Venus, Saturn, and Jupiter.  For all of the data, the pointing varied by less than 4$''$.  These maps were corrected for the effects of atmospheric opacity using ``skydip'' scans taken prior to beginning and upon completion of each map.  This correction is accomplished by fitting a polynomial function to the optical depth, at 870\,\micron\, as a function of elevation.  The bolometer data were flux calibrated with images of Mars, Saturn, and the known submillimeter secondary calibrator source IRAS 16293-2422 \citep{san94} taken using a continuum on/off sequence technique.  We estimate the flux calibration to be accurate within the $\sim$\,20\% level. 

\subsection{Spectral line observations}

\subsubsection{HHT data}

In February and May of 2004, we obtained $^{12}$CO$\,(3-2)$ and $^{13}$CO$\,(3-2)$ spectral line data with the HHT toward known outflow regions, NGC~2264~O and NGC~2264~H, associated with IRAS~25 and 27, respectively.  A summary of important observational parameters for these data is presented in Table~\ref{tab:observations}.  A set of $300'' \times 300''$ $^{12}$CO\,$(3-2)$ fully sampled OTF maps were made surrounding the two IRAS sources.  The pointing was monitored similar to the continuum observations and varied $<5$$''$.  A total of two separate maps, each with two polarization data, were made for NGC~2264~O \& H.  After careful calibration, maps were combined to produce the data described in Table~\ref{tab:observations}.  In addition, $^{13}$CO\,$(3-2)$ Absolute Position Switched (APS) spectra were taken toward the locations of the IRAS sources as published in \cite{mly89}.  A total integration time of 140 s was spent on source for each APS $^{13}$CO observation.  The $2\sigma$ RMS of all observations has been included in Table~\ref{tab:observations}.

The HHT facility dual-polarization, double-sideband 345\,GHz receiver was used with two Acousto-Optical Spectrometer backends, each possessing 1\,GHz total bandwidth and 2048 channels.  Individual APS spectra were taken toward the positions of IRAS~25 and 27 at the beginning of each observing session and used to cross calibrate data sets taken on different days.  Orion A ($\alpha_{1950}=05^{h} \, 32^{m} \, 47.0^{s},~\delta_{1950}=-05^{\circ} \, 24' \, 21''$) was used to calibrate the dual polarization data before they were combined.  For ease of comparing data sets from different telescopes, the main beam efficiency ($\eta_{mb}$) of the 345\,GHz receiver was measured and used to convert antenna temperature corrected for atmospheric extinction, $T_{A}^{*}$, to a main beam brightness temperature, $T_{mb}$, according to $T_{mb}=T_{A}^{*}/\eta_{mb}$.  Measurements of Saturn were used to find $\eta_{mb}$ using Equation~\ref{eq:etamb}, where $T_{R}(planet)$ is the planet's brightness temperature, D is the angular diameter of the planet when it was observed, and $\theta$ is the telescope beam size.  Measurements of Saturn made during observations resulted in $\eta_{mb}$ values that varied $\sim$\,20\%.  Based on this variation, we adopt a value of $\eta_{mb}\approx 60\%$ for the 345\,GHz receiver during this time.            

\begin{equation}
\label{eq:etamb}
\eta_{mb}=\frac{T_{A}^{*}(planet)}{T_R(planet)}{\lbrack{1-exp(\frac{-D^2}{\theta^2}{ln2})}\rbrack}^{-1}
\end{equation}

\subsubsection{12m data}

APS $^{12}$CO\,$(1-0)$ observations of IRAS~25 and 27 were made at the ARO 12m millimeter-wave telescope near Tucson, AZ during April 2002.  A summary of the observational information is presented in Table~\ref{tab:observations}.  A total integration time of 120~s was spent on source for each observation and $2\sigma$ RMS values are listed in Table~\ref{tab:observations}.  The 12m dual-polarization, single-sideband $90-116$\,GHz receiver and Millimeter Autocorrelator (MAC) backend with 300\,kHz total useable bandwidth and 8192 channels were used.  Similar to the HHT data, both polarizations were combined in order to improve the signal-to-noise using Orion A as a cross-calibrator.  The 12m makes measurements in terms of $T_{R}^{*}$, the observed antenna temperature corrected for atmospheric attenuation, radiative loss, and rearward scattering and spillover.  In order to make meaningful comparisons between data sets, these measurements were converted to a $T_{mb}$ scale using a procedure documented in Appendix C of the 12m manual.  According to the manual, $T_{mb}=T_{R}^{*}/1.08$ for the $90-116$\,GHz receiver at 115\,GHz.  With this conversion, the 12m data were compared with HHT spectra after the data sets were convolved to the same beam size.  Measurements of $T_{mb}$ for the 12m data are accurate to within $\sim$\,20\%.     

\section{Results}

\subsection{Continuum Analysis}

Results of the $870\, \micron$ HHT continuum data analysis are presented for IRAS~25 \& 27.  Table~\ref{tab:870srcflux} lists the fluxes found at the positions of sources IRAS~25 \& 27 and the associated continuum cores.  The cores, labeled with S1, S2, and S3 extensions, were initially identified as emission peaks in $850\, \micron$ and $450\, \micron$ SCUBA maps \citep{wc03}.  We derived $870\, \micron$ peak fluxes within multiple apertures for each of the sources, including 45$''$ (for easy comparison with \cite{wc03}) and 3$'$$\times$5$'$, the size of the IRAS error ellipse at 100 \micron.   

\begin{figure}
\includegraphics[width=3.2in]{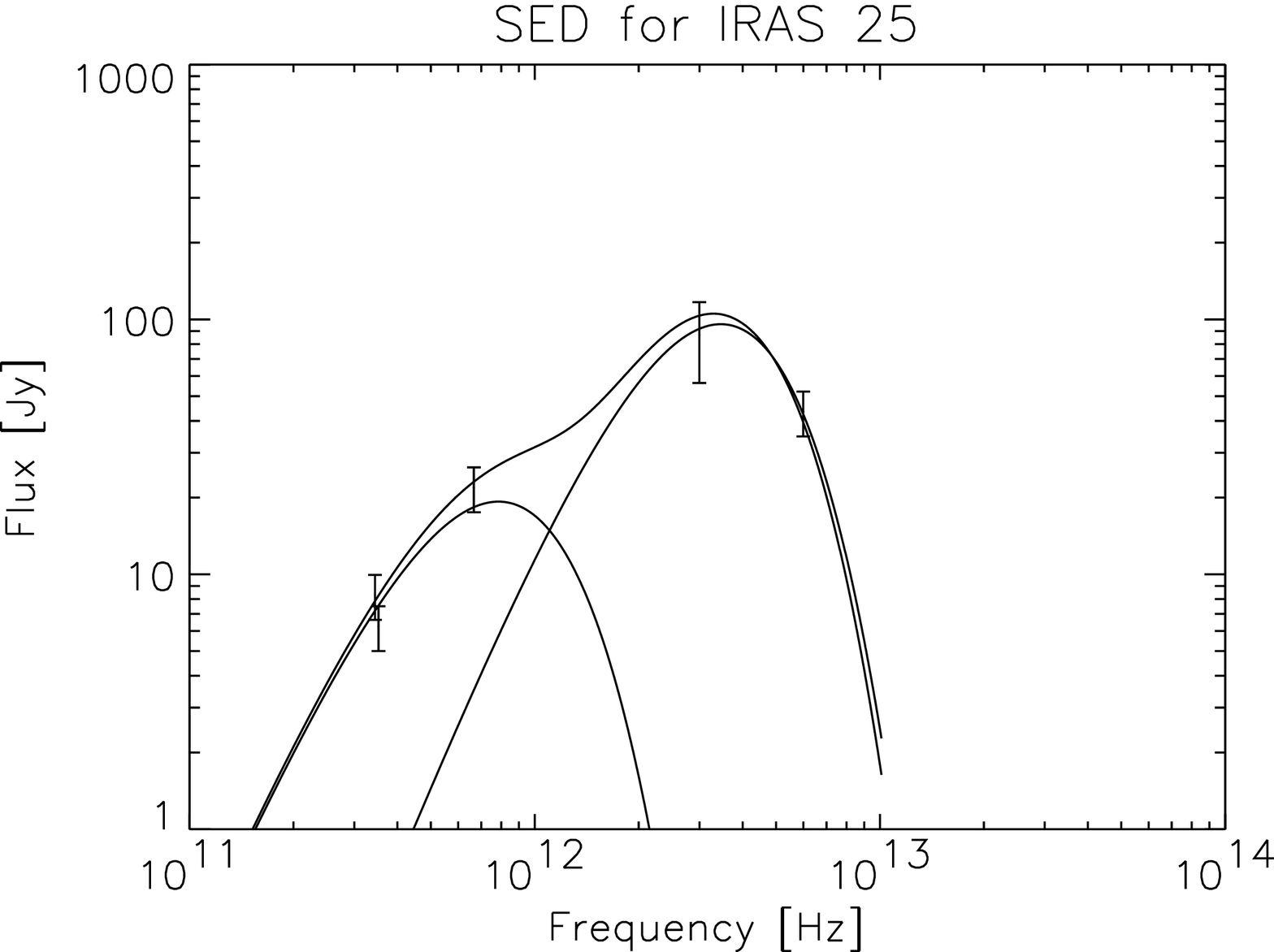}
\includegraphics[width=3.2in]{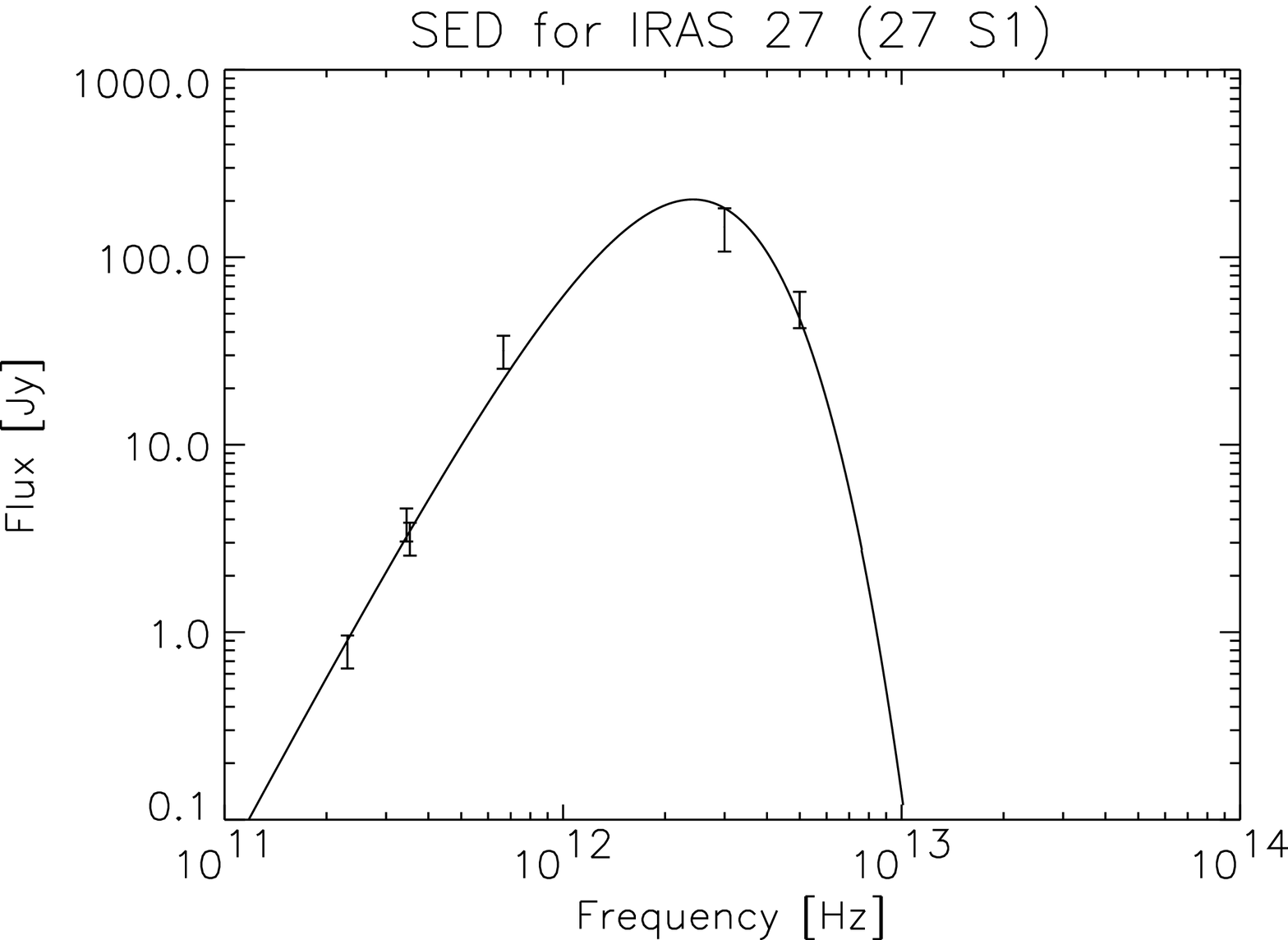}
\includegraphics[width=3.2in]{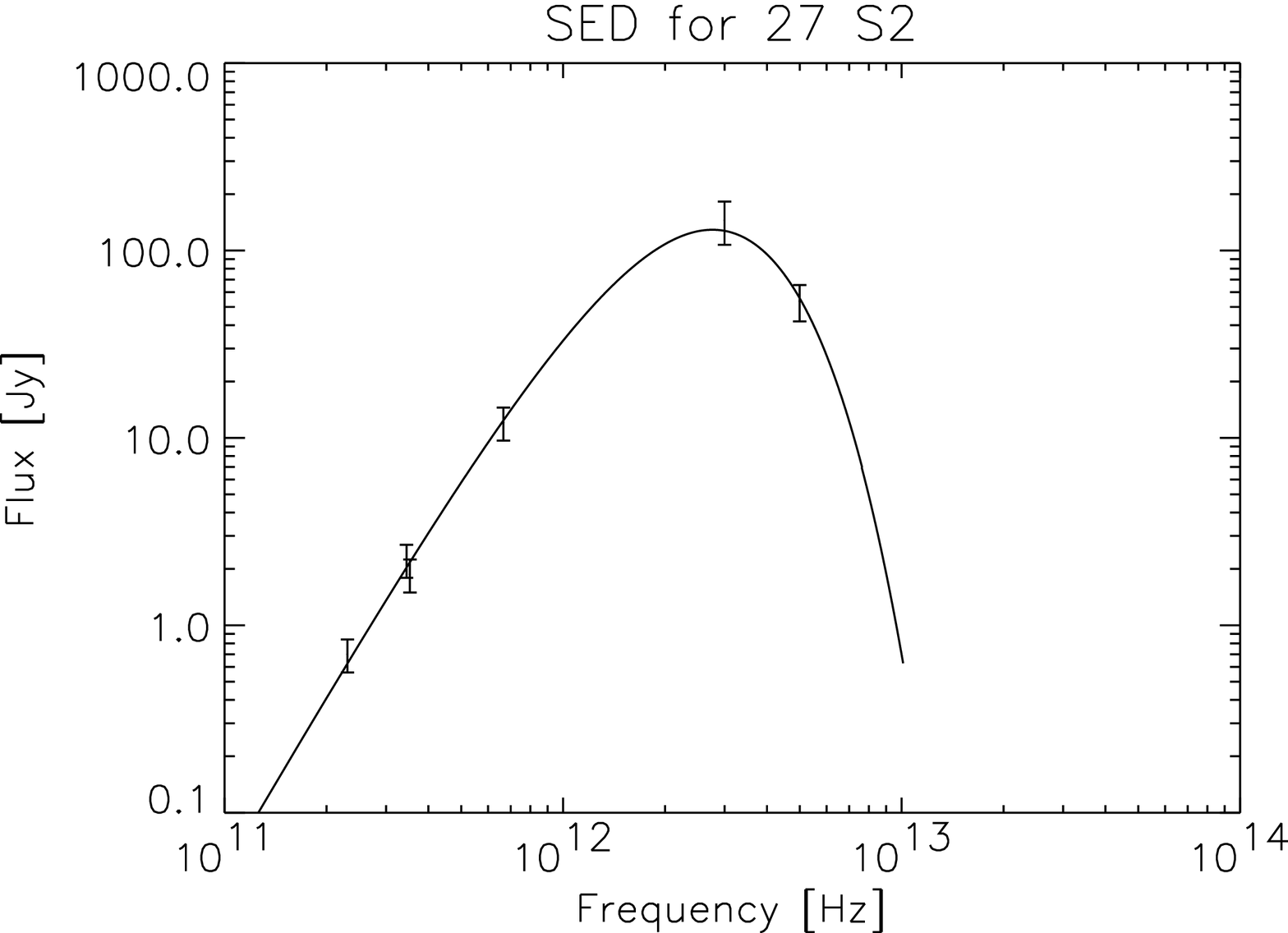}
\includegraphics[width=3.2in]{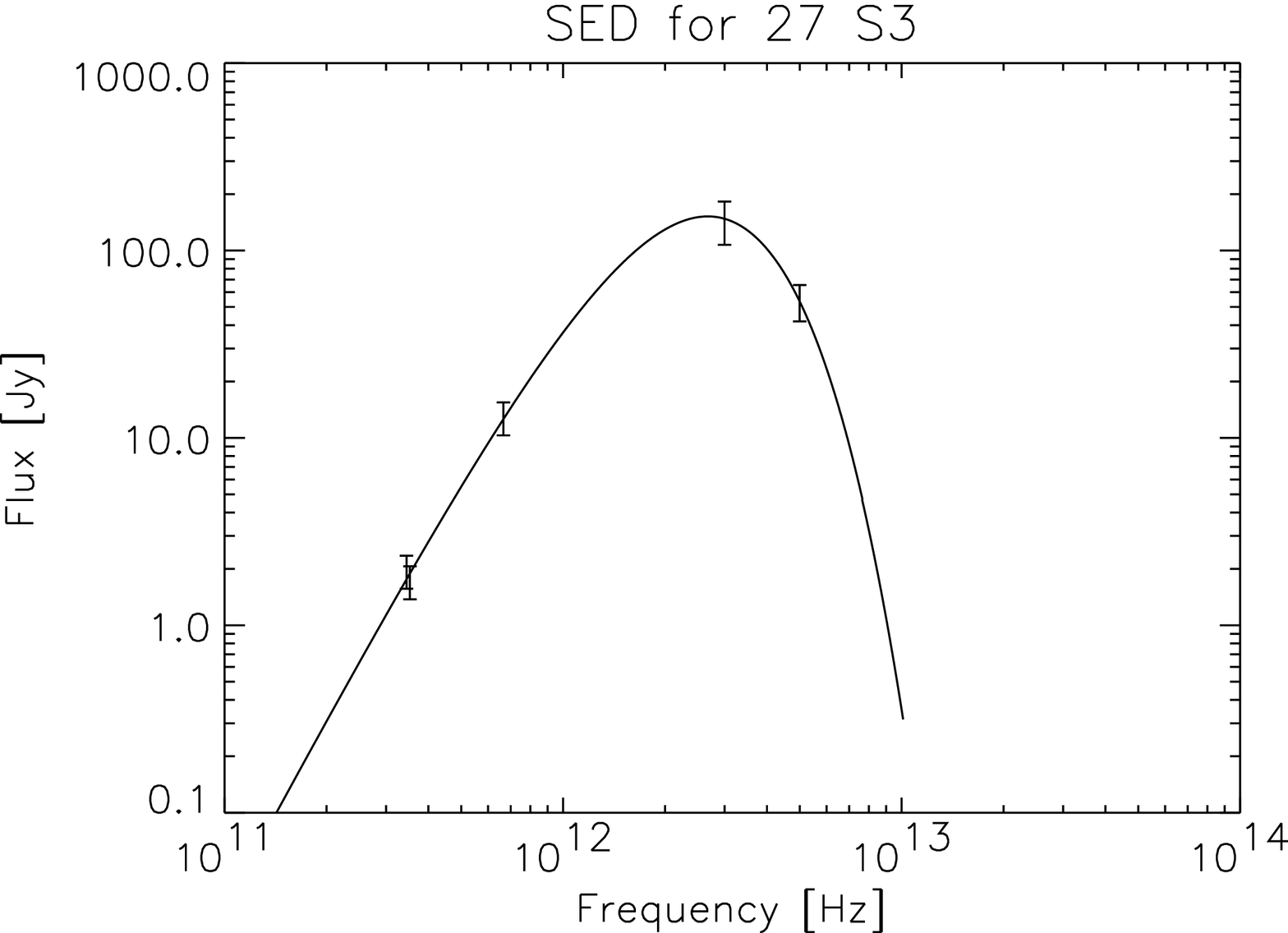}
\caption{\label{fig1} FIR SEDs for sources IRAS~25, IRAS~27 (27~S1), 27~S3, \& 27~S2 (clockwise from upper left).  From the shortest wavelength, fluxes included are IRAS 60$ \, \micron$ and 100$ \, \micron$ data, $450\,\micron$ and 850$ \, \micron$ SCUBA measurements, 870$ \, \micron$ HHT data from this study, and where available (IRAS~27 and 27~S2), 1300$ \, \micron$ fluxes.  For IRAS~25, a two component BB function was fit to this SED.  All other sources are well described by single temperature functions.}
\end{figure}

Combining our $870\, \micron$ continuum data with bolometric observations of other observers and IRAS, we construct FIR SEDs (see Figure~\ref{fig1}) for dust cores with 870\,\micron~fluxes presented in Table~\ref{tab:870srcflux}.  When IRAS data were used, the fluxes were taken from the PSC.  At 60\,\micron~\& 100\,\micron, the IRAS the beam sizes are larger than the IRAS~25 source, indicating fluxes of the entire clump emission were found.  When constructing the IRAS~25 FIR SED, measurements of the $870\,\micron$~emission of the whole clump were used.  In order to incorporate SCUBA $450 \, \micron\,\&\,850 \, \micron$ data, we estimated the total flux found at these wavelengths by summing published fluxes within 45$''$ apertures of the 25 S1 and S2 cores.  For IRAS~27, 60\,\micron~\& 100\,\micron~IRAS fluxes were incorporated in the FIR SED with the $870 \, \micron$ fluxes measured within 45$''$ apertures around the continuum cores IRAS~27 (27 S1), S2, and S3.  Published SCUBA $450 \, \micron~\&~850 \, \micron$ measurements made within the same 45$''$ apertures were included.  When available (for sources 27 S1 \& S2) $1300 \, \micron$ fluxes were incorporated.  A summary of continuum measurements for each source at all wavelengths where data are available is in Table~\ref{tab:srcflux} along with associated references.

In order to probe physical conditions of the continuum cores, FIR SEDs were fit with expressions of the following form \citep{hil83}:
\begin{equation}
\label{eq:Fnu}
F_{\nu}=\Omega_{s}B_{\nu}(T)(1-e^{\tau_{\nu}})
\end{equation}
\noindent
where $\Omega_{s}$ is the source size and
\begin{equation}
\label{eq:PlanckTau}
B_{\nu}(T)=\frac{2h\nu^{3}/c^{2}}{(e^{h\nu/kT}-1)}~~~~~\&~~~~~\tau_{\nu}=(\nu/\nu_{o})^{\beta}
\end{equation}
\noindent
In Equation~\ref{eq:PlanckTau}, $B_{\nu}(T)$ is the Planck function, $\tau_{\nu}$ describes the frequency dependent optical depth, $\beta$ is the spectral index governing dust grain emissivity properties,  and $\nu_{o}$ is the frequency where $\tau_{\nu} = 1$.  

The results of fitting flux values to Equations~\ref{eq:Fnu} and \ref{eq:PlanckTau} are illustrated in Figure~\ref{fig1} and Table~\ref{tab:sedparms} summarizes the details numerically.  We have only attempted to fit the cold dust component ($\lambda \ge 60 \, \micron$) of the continuum emission that is responsible for the majority of the flux at mm and submm wavelengths.  In all cases, except for IRAS~25, each continuum core was well fit by a single temperature blackbody (BB), similar to Equation~\ref{eq:PlanckTau}.  In order to obtain a good least-squares fit to each of these dust cores (IRAS~27 (27 S1), 27 S2, and 27 S3), source size ($\Omega_{s}$), temperature (T), and spectral index ($\beta$) parameters were varied iteratively until a best fit solution was found.  

Fitting SED curves to IRAS~25 flux measurements proved more complex, and a two temperature component BB function was used.  In order to obtain a good, physically plausible fit to these data, both dust temperature components, T, and $\beta$ indices were selected and held constant while the source sizes, $\Omega_{s}$, were iteratively fit.  The goodness of fit did not strongly depend upon the values of $\beta$ chosen for the two components and therefore the resulting fit does not reflect unique or precise values of spectral index.  Since spectral index is probably constant within the same molecular cloud complex, our choice of $\beta$ used to fit the IRAS~25 emission was guided by solutions from iterative fits of IRAS~27 (27 S1), 27 S2, and 27 S3 SEDs.  The temperatures of the two components fit to IRAS~25, 10\,K and 40\,K, were selected to be similar to the $\sim$\,30\,K obtained from using Wien's displacement law to estimate the characteristic dust temperature for IRAS~25.  The two component BB fit was found to be rather sensitive to the values of temperature chosen.  Based upon this sensitivity, the temperatures of the components are probably accurate to within 5K.    

Luminosities ($L_{FIR}$) of continuum cores were found using $L_{FIR}=3.1 \times 10^{-10} D^{2} \int F_{\nu} d\nu$ \citep{wal90} where $L_{FIR}$ is in L$_{\odot}$, $D$ is in pc, $F_{\nu}$ is in Jy, and $\nu$ is in GHz.  Dust filling factors, $f_{d}=\Omega_{s} / \Omega_{m}$, were estimated by comparing values of $\Omega_{s}$ derived from the SEDs to half-power $870\,\micron$~source sizes, $\Omega_{m}$.  Table~\ref{tab:sedparms} presents the results of the FIR SED model fits and parameters for each of the cores examined.  Using techniques similar to \cite{hil83}, physical properties of the dust cores, including dust optical depth at $870\,\micron$~($\tau_{\nu}=\tau_{870}$), column density ($N_{H_{2}}$), and total gas mass ($M_{H_{2}}$) were derived.  Number density values ($n_{H_{2}}$) were found using the relation $M_{H_{2}}/(4/3\,\pi\,R^{3}\,m_{H_{2}})$ where $m_{H_{2}}$ is the mass of a hydrogen molecule and $R$ is the radius over which $M_{H_2}$ was found.  The visual extinction was estimated using the relation $A_{v} = 1.1 \times 10^{-21} N_{H_{2}}$ \citep{bsd78}.  The results of this analysis are presented in Table~\ref{tab:coreproperties}.  In making these computations, we assumed a gas to dust ratio of 100 in mass, a grain density of $\rho=3$ g cm$^{-3}$, and a dust particle radius of $a=0.1 \, \micron$.  These values are consistent with those determined by Hildebrand and are commonly used for molecular clouds in the galaxy.  

At $870\,\micron$, Figure \ref{fig1} shows that the majority of the flux observed for IRAS~25 is due to the colder 10\,K BB component.  It is reasonable to obtain an upper limit to the amount of mass in the IRAS~25 core region by assuming that all flux at these wavelengths is attributable to the cold gas.  Using the analysis tools described above, a limit of 140\,M$_{\odot}$ was obtained for the IRAS~25 core (Table \ref{tab:coreproperties}).  Although this mass limit is large compared with the IRAS~27 core, it is not unexpected.  The observed flux at $870\,\micron$ for IRAS~27 (Figure \ref{fig1}) is well described by a 30\,K BB source and is weaker by about a factor of two when compared with IRAS~25 $870\,\micron$ fluxes (Table \ref{tab:srcflux}).  At these wavelengths, we observe comparatively more flux from a colder source, implying larger column density and core mass values for IRAS~25.   

\subsection{Spectral Line Observations}

\begin{figure}
\includegraphics[scale=0.35]{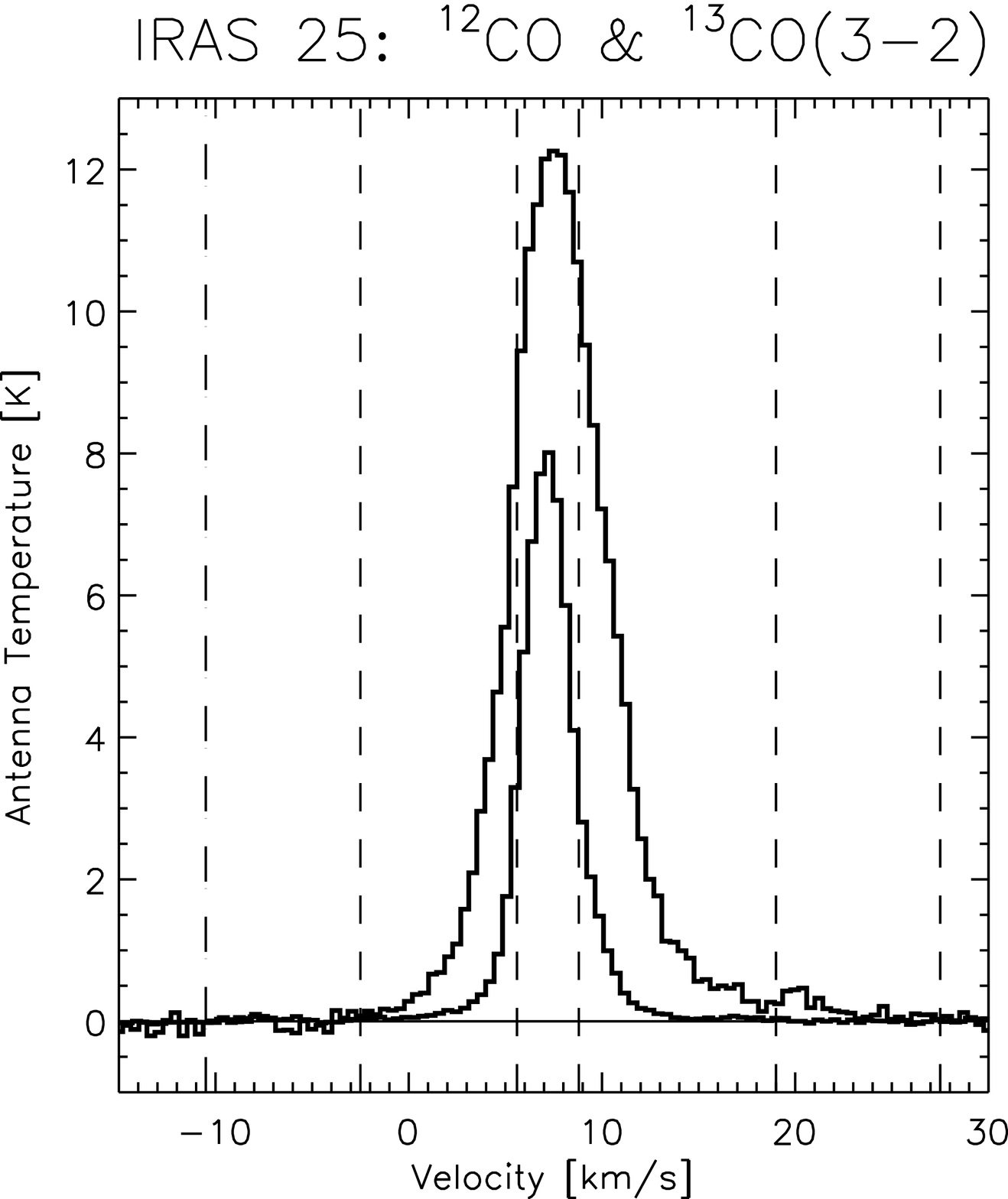}
\includegraphics[scale=0.35]{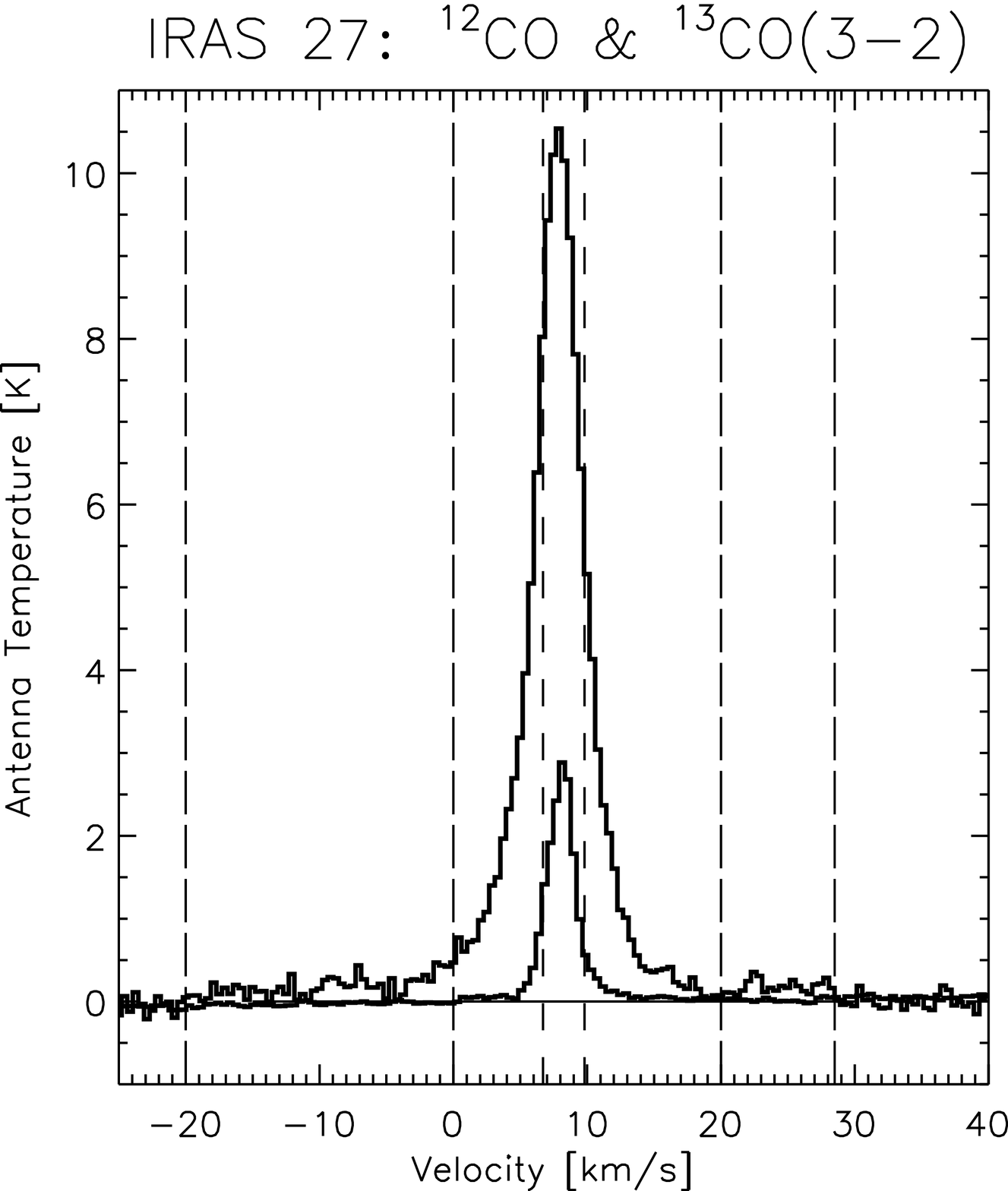}
\caption{\label{fig2} $^{12}$CO $\&~^{13}$CO\,$(3-2)$ overlayed spectra of IRAS~25 and 27.  Vertical lines show LV and HV red and blue wings and line core (see Table~\ref{tab:linewings}).}
\end{figure}

In this section, the results of $^{12}$CO\,$(3-2)$ $\&~^{13}$CO\,$(3-2)$ analyses are presented for IRAS~25 \& 27.  The CO$(3-2)$ spectra for both objects are single peaked features and possess broad line wings ($\sim$\,40 km s$^{-1}$ in extent for IRAS~25 and $\sim$\,50 km s$^{-1}$ for IRAS~27).  In order to investigate the spatial distribution of the outflow molecular gas, emission line wings were defined by comparing the $^{12}$CO\,$(3-2)$ $\&~^{13}$CO\,$(3-2)$ spectra (see Figure~\ref{fig2}) and then used to make contour maps.  The $^{13}$CO\,$(3-2)$ line wings set the maximum extent of low velocity emission associated with the outflows.  High velocity wings extended to where the $^{12}$CO\,$(3-2)$ emission merged with the noise.  Figure~\ref{fig2} illustrates these regions, including the line core, low, and high velocity wings for both IRAS~25 and 27 and Table~\ref{tab:linewings} summarizes this information.  In this table, the spectral line emission of the entire (full) red and blue wings (F~RW \& F~BW) has been divided into low and high velocity (LV \& HV) regions.   

\subsubsection{IRAS~25 (IRAS 06382+1017)}              
\begin{figure}
\includegraphics[scale=0.35]{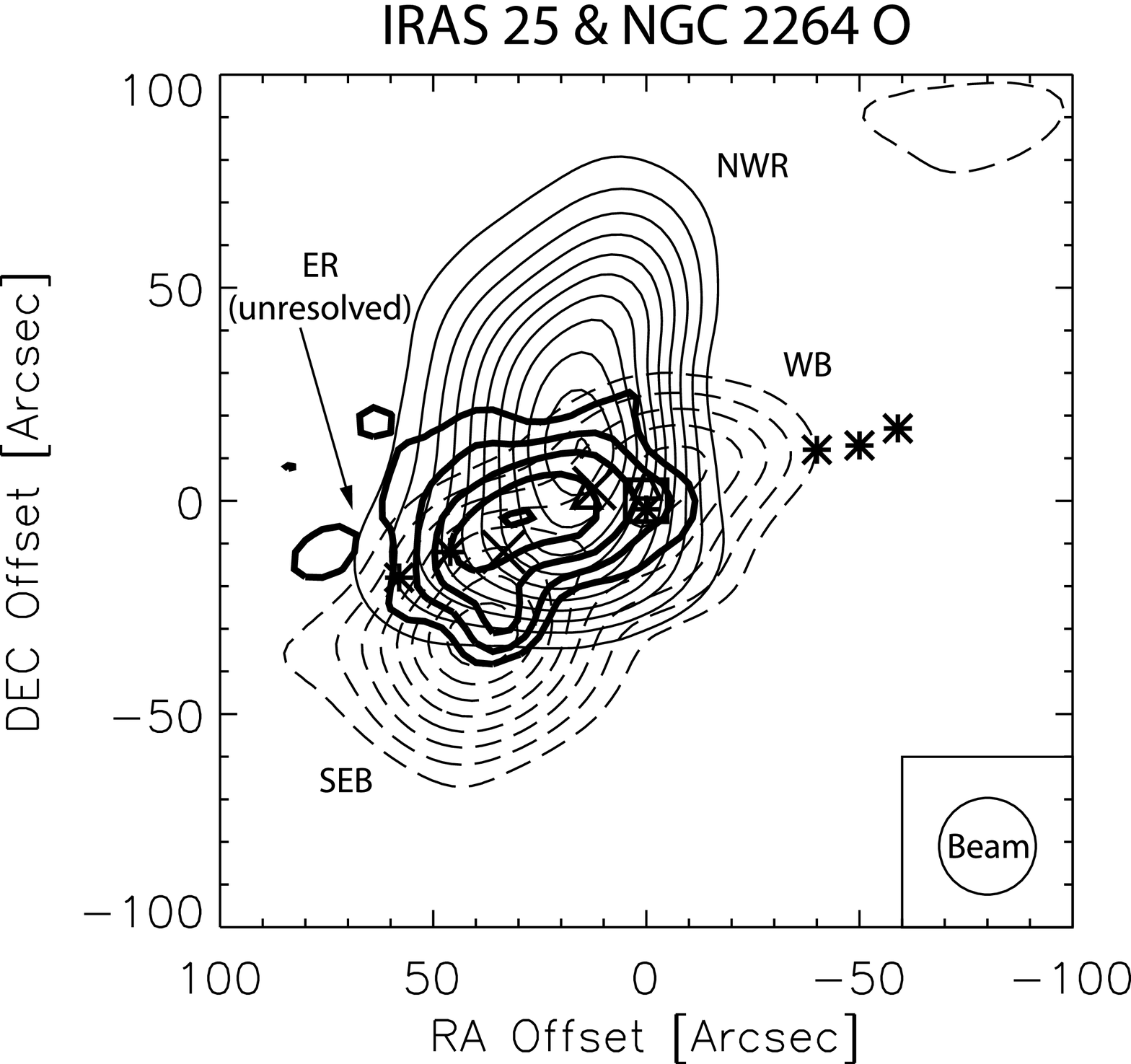}
\includegraphics[scale=0.35]{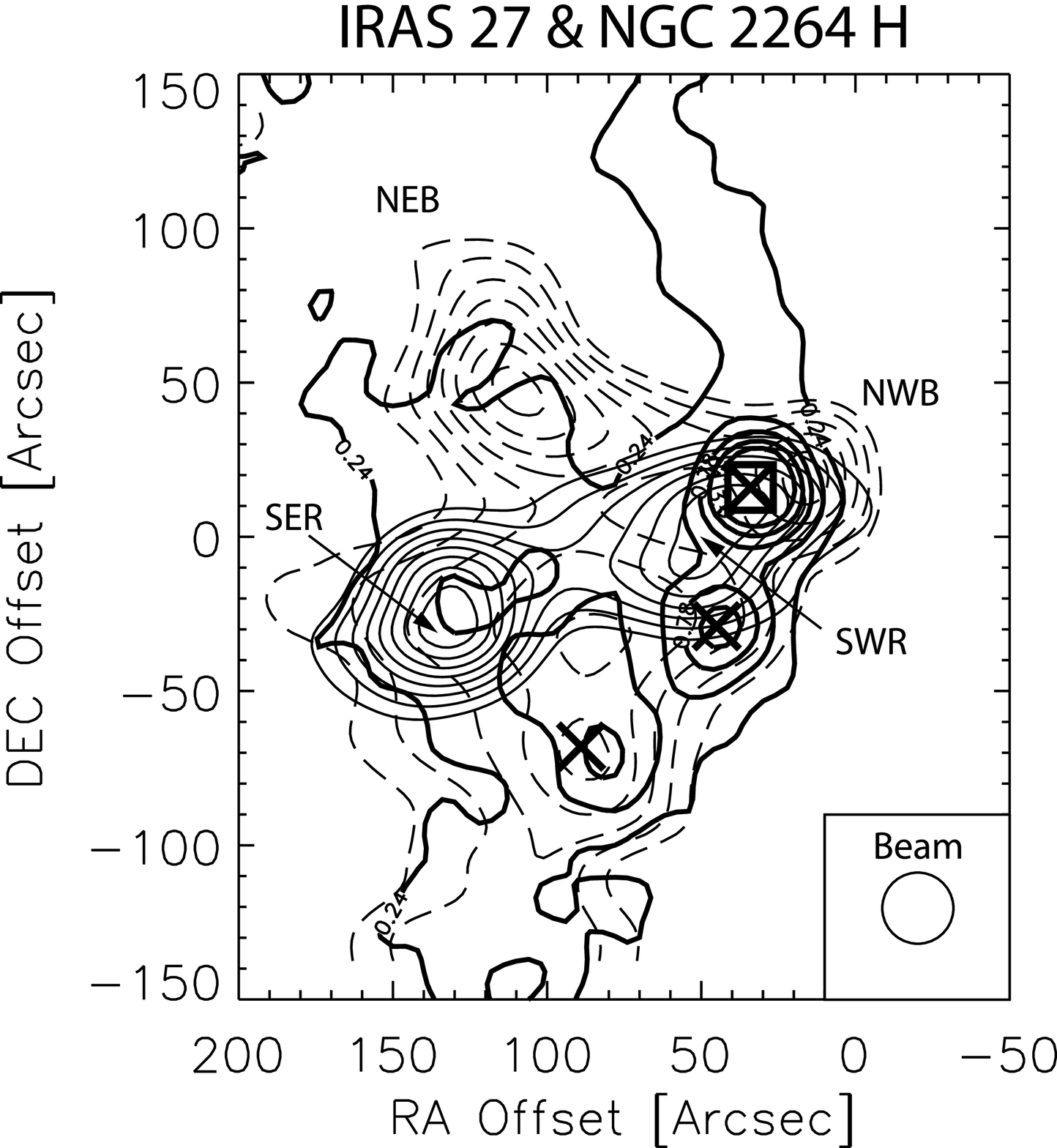}
\caption{\label{fig3} Contours of red (solid) and blue (dashed) $^{12}$CO\,$(3-2)$ outflow emission associated with NGC~2264~O \& H overlayed on 870$\,\micron$ continuum maps (thick solid lines).  Locations of sources within each region are shown, including IRAS~25 \& 27 (squares), submillimeter continuum objects (``X''), HH~124 emission knots (asterisks), and VLA sources (triangles).  Continuum $870 \, \micron$ contours begin at $2\sigma=0.38$\,Jy and $3\sigma=0.16$\,Jy for IRAS~25 \& 27, respectively, increasing in steps of 0.29\,K \& 0.27\,K toward emission peaks, 1.85\,Jy and 2.25\,Jy.  F~RW contours begin at 6.6\,K\,km\,s$^{-1}$ for NGC~2264~O \& H increasing in 2.6\,K\,km\,s$^{-1}$ steps toward emission peaks, 36\,K\,km\,s$^{-1}$ and 31\,K\,km\,s$^{-1}$, respectively.  F~BW contours start at 6.3\,K\,km\,s$^{-1}$ and 6.8\,K\,km\,s$^{-1}$ for NGC~2264~O \& H, respectively, and increase by 1.3\,K\,km\,s$^{-1}$ and 2.6\,K\,km\,s$^{-1}$.  Peak F~BW emission is 17\,K\,km\,s$^{-1}$ and 32\,K\,km\,s$^{-1}$ for NGC~2264~O \& H, respectively.}
\end{figure}

\begin{figure}
\includegraphics[scale=0.35]{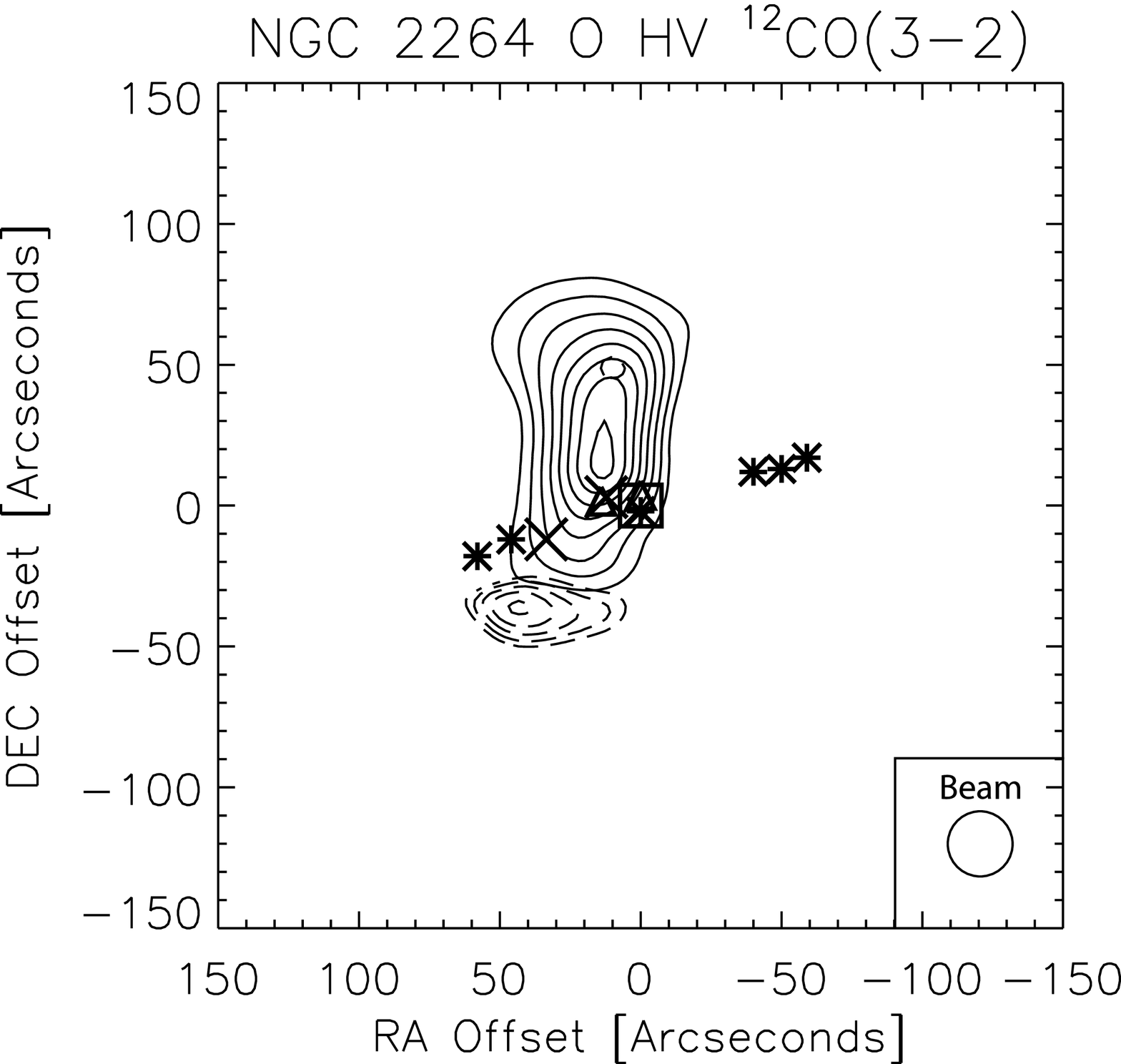}
\includegraphics[scale=0.35]{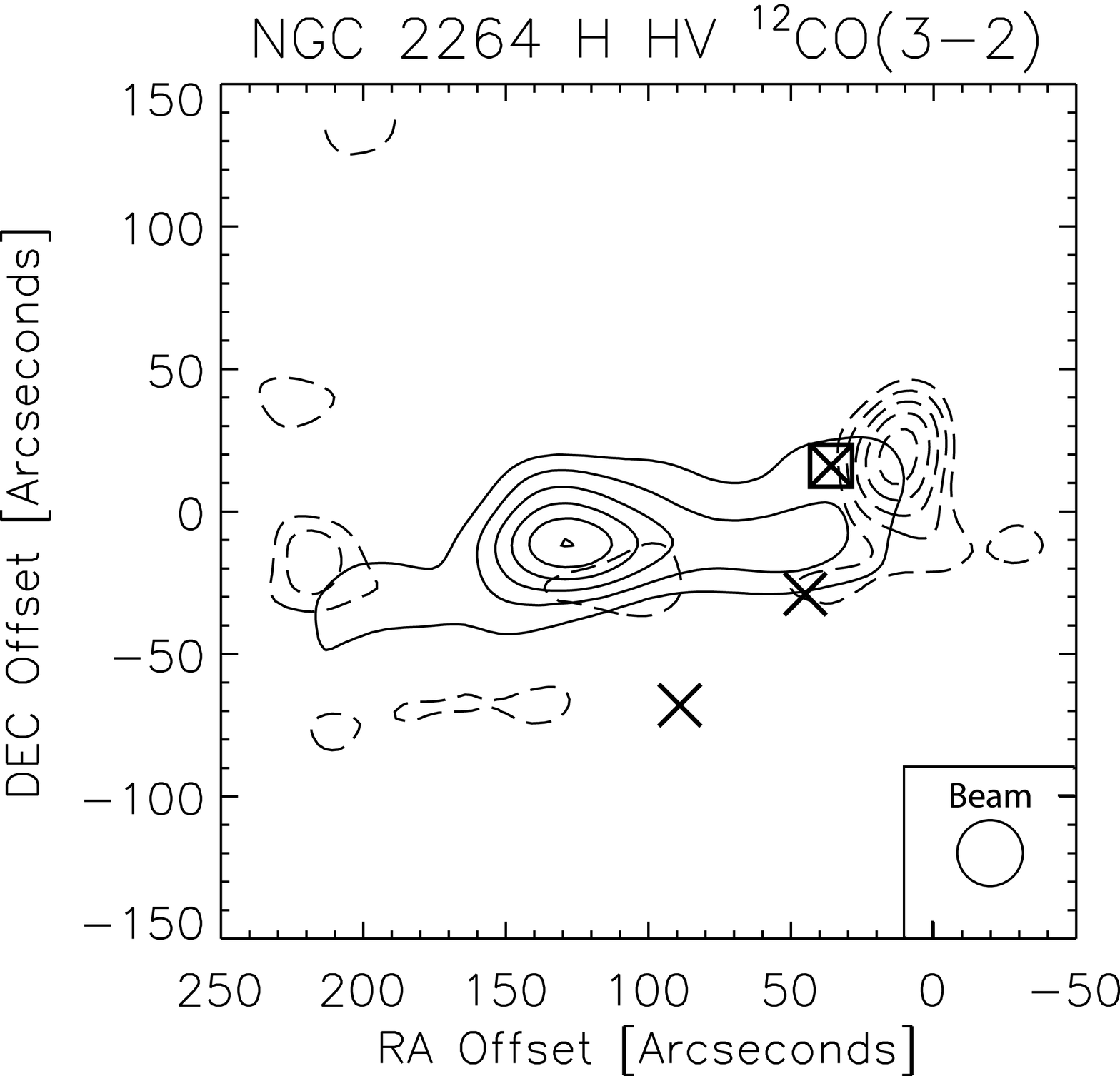}
\caption{\label{fig4} Contours of the HV red (solid) \& blue (dashed) $^{12}$CO\,$(3-2)$ emission (see Table~\ref{tab:linewings}) associated with NGC~2264~O \& H.  HV blue contours begin at $2\,\sigma=1.6$\,K km s$^{-1}$ and $2\,\sigma=1.5$\,K km s$^{-1}$, increasing in steps of 0.15\,K km s$^{-1}$ and 0.7\,K km s$^{-1}$ for NGC~2264~O \& H, respectively.  Contours of HV red emission begin at $2\,\sigma=1.8$\,K km s$^{-1}$ and $2\,\sigma=1.0$\,K km s$^{-1}$, increasing by 0.8\,K km s$^{-1}$ and 1.0\,K km s$^{-1}$.  HV emission peaks at 2.2\,K km s$^{-1}$ and 5.4\,K km s$^{-1}$ (blue) and 8.2\,K km s$^{-1}$ and 7.0\,K km s$^{-1}$ (red) for NGC~2264~O \& H, respectively.}
\end{figure}

Contour maps of F~RW and F~BW integrated intensity were constructed for IRAS~25 using $I_{F~RW}=\int_{8.8~ km\,s^{-1}}^{27.5~ km\,s^{-1}}T_{A}^{*~12}dv$ \& $I_{F~BW}=\int_{-10.5~ km\,s^{-1}}^{5.6~ km\,s^{-1}}T_{A}^{*~12}dv$, respectively.  Figure \ref{fig3} shows red (thin, solid contours) and blue wing (thin, dashed lines) $^{12}$CO\,$(3-2)$ emission associated with outflow NGC~2264~O surrounding IRAS~25.  Contours are superimposed on an $870\,\micron$ continuum OTF map (see Section (3.1)) and the figure includes locations of other sources discovered within the same region.  Figure \ref{fig3} shows two prominent blue outflow lobes, SEB \& WB, and an elongated red emission lobe.  Although a single red outflow lobe is detected, observations indicate that the emission is unresolved and two separate red lobes, ER \& NWR, exist.  Figure~\ref{fig4} is a contour map of the high velocity (HV) red and blue $^{12}$CO\,$(3-2)$ emission associated with NGC~2264~O.  A large fraction of the eastern red emission (ER) disappeared for the high velocity gas, leaving a clear view of the compact northwestern red (NWR) emission.  This indicates that the ER and NWR emission possess different velocity components and suggests that the emission belongs to two different outflow lobes.  

Our data support the existence of two $^{12}$CO\,$(3-2)$ outflows in the region.  Figure~\ref{fig4} aided in determining the orientations of the two outflows associated with NGC~2264~O.  The high velocity gas traces an outflow oriented southeast (SEB) to northwest (NWR).  The WB lobe and a portion of ER emission do not appear in this figure.  This indicates these lobes have similar velocities and suggests they describe a second outflow.  The existence of multiple molecular outflows within this region has been suggested by other authors \citep{php95}, \citep{wc03}.  Our observations of two $^{12}$CO\,$(3-2)$ outflows fit well with the work of \cite{rei04} and lend support to observations of recurring outflow activity taking place in the region surrounding IRAS~25 (\cite{ogu95}, \cite{wc03}, \cite{rei04}).

\begin{figure}
\includegraphics[scale=0.75]{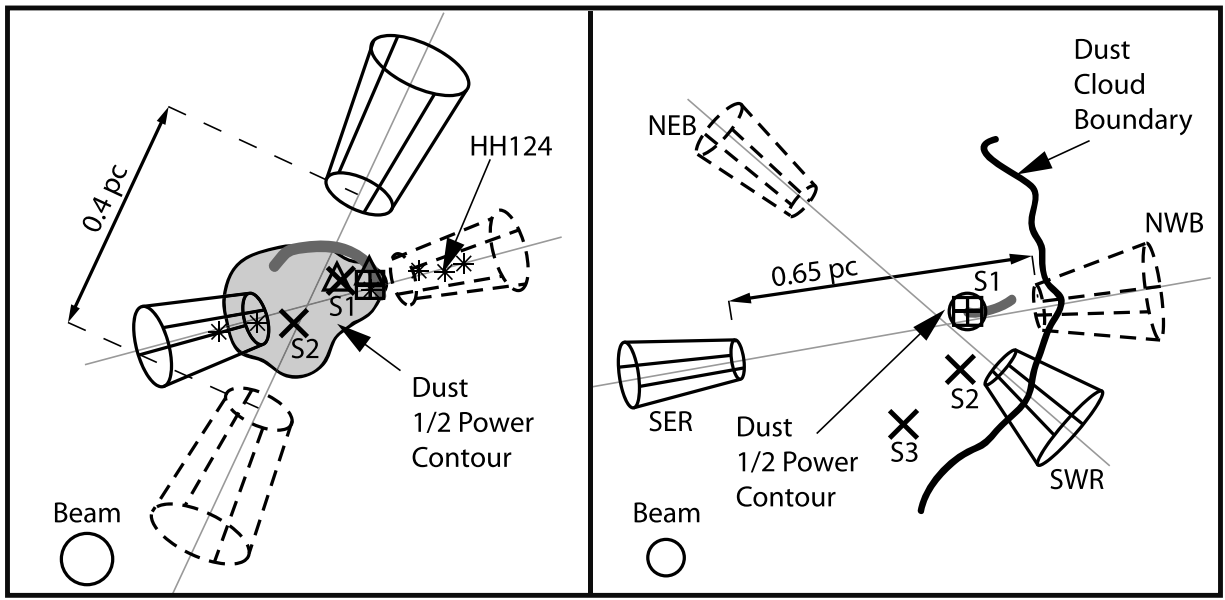}
\caption{\label{fig5} Model of the $^{12}$CO\,$(3-2)$ outflows surrounding IRAS~25 \& 27 (left to right).  The red (solid) \& blue (dashed) lobes of the outflows and their approximate positions with respect to half power $870\,\micron$ continuum contours are indicated.  Two molecular outflows are present in each region.  The sizes, orientations, and locations of IR reflection nebulae associated with IRAS~25 \& 27 are indicated with solid arcs.}
\end{figure}

For clarity, the information contained in Figure~\ref{fig3} is summerized in the model sketch of the IRAS~25 region shown in Figure~\ref{fig5}.  The location of IRAS 25 is denoted with a box, and continuum peaks 25 S1 \& 25 S2 associated with $850\, \micron \, \& \,450\, \micron$ SCUBA emission are shown with ``X'' symbols.  IRAS~25 has been identified as the driving source for a large scale (0.48\,pc) Herbig-Haro flow, HH~124 \citep{wor92}.  Six knots of emission were observed (asterisks) and lie along the WB$-$ER outflow axis.  This region also includes an IR reflection nebula (thick, gray contour in Figure \ref{fig5}) associated with IRAS 06382+1017 \citep{php95} and two VLA sources (triangles) within the IRAS error ellipse \citep{rr98}.           

In this complex environment, it is difficult to determine sources of the $^{12}$CO\,$(3-2)$ outflows.  Our observations and the work of \cite{rei04} indicate that the WB$-$ER outflow axis is well aligned with HH~124, suggesting that the IRAS object may be the source of this outflow emission.  The axis of the SEB$-$NWR outflow is in proximity to many objects, including IRAS~25 and the submillimeter continuum sources.  The source of this molecular outflow has not been determined \citep{rei04}, although it is oriented along the opening angle of the IR reflection nebula associated with IRAS~25 (see Figure~\ref{fig5}).  It is possible that two $^{12}$CO\,$(3-2)$ molecular outflows emanate from IRAS~25; the WB$-$ER flow oriented along HH~124, and a SEB$-$NWR outflow along the IR reflection nebula opening angle. 

The SEB$-$NWR outflow axis does not intersect well with IRAS~25.  If the IRAS object is the source of the flow, this may imply that the outflow is redirected as it encounters dense ambient material.  Figure~\ref{fig3} shows that the IRAS source is located near the edge of a dense, extended $870\,\micron$ continuum core.  We estimate how dense this material would have to be in order to redirect the outflow and make comparisons with the results of the continuum analysis ($n_{H_2}$ in Table~\ref{tab:coreproperties}).  To redirect the flow, the clump pressure, given by the ideal gas law ($n_{cl}\,k\,T$) in LTE, must exceed the pressure of the outflowing material, $\dot{P}_{OF}/(\pi\,R^{2})$.  For an order of magnitude estimate, it is assumed the outflow acts mostly over an area encompassing its 1/2 power contours.  In this analysis, $k$ is Boltzmann's constant, $T = 10$\,K is the core temperature derived from continuum observations (see Table~\ref{tab:sedparms}), $\dot{P}_{OF}$ is the outflow momentum rate (see Section (3.2.3)), and $R$ is the average radius of 1/2 power outflow contours.  The minimum core density capable of redirecting the outflow is $n_{cl} \sim 10^{7}$\,cm$^{-3}$.  This is large compared with $n_{H_2}$ (see Table~\ref{tab:coreproperties}), suggesting the density of the core is most likely not sufficient to influence outflows in the region.
 
It was suggested that IRAS~25 is a barely resolved binary system with a possible component $\sim$\,1400 AU (1.8$''$ at the assumed distance of NGC~2264) from the primary at a position angle of $\sim155^{\circ}$ \citep{php95}.  The presence of two outflows is consistant with IRAS~25 being a binary system and may explain the unusual morphology.  The $870\,\micron$ continuum emission (Section (3.1)) can be used to estimate the mass of stellar objects in the system.  Assuming that the bolometric luminosity ($L_{FIR}$) of this object (see Table~\ref{tab:sedparms}) equals the accretion luminosity ($L_{acc} = G M \dot{M}_{acc} / R$), an upper limit to the mass of the central object is obtained.  We assumed a typical radius of $R = 2 \times 10^{11}$ cm, $\sim$\,3 R$_{\odot}$, for the hydrostatic core of an accreting protostellar object \citep{sst80} and the mass infall rate of an isothermal cloud, $\dot{M}_{acc} = 0.975\, c_{s}^{3} / G$ \citep{shu77}, where $c_{s} = \sqrt{k\, T / m_{H_{2}}}$ is the cloud sound speed.  Since material is accreted from a reservoir in close proximity to the protostar, the warmer continuum component $T = 40$\,K (see Table~\ref{tab:sedparms}) was used to calculate $c_{s}$.  This simple analysis estimates 0.5\,M$_{\odot}$ for the stellar object(s) associated with IRAS~25.         

\subsubsection{IRAS~27 (IRAS 06381+1039)}
       
Similar to (Section (3.2.1)), $^{12}$CO\,$(3-2)$ line wing integrated intensities were found for NGC~2264~H outflows surrounding IRAS~27 using $I_{F~RW}=\int_{9.8~ km\,s^{-1}}^{28.5~ km\,s^{-1}}T_{A}^{*~12}dv$ \& $I_{F~BW}=\int_{-20.0~ km\,s^{-1}}^{6.7~ km\,s^{-1}}T_{A}^{*~12}dv$.  Figure~\ref{fig3} shows the outflow emission (F~RW\,$=$\,solid, thin lines and F~BW\,$=$\,thin, dashed contours) superimposed on a map of $870 \, \micron$ continuum emission.  Other sources in the region are shown, including the location of the IRAS source (box) and $870~\&~450 \, \micron$ SCUBA emission peaks, 27 S1, 27 S2, \& 27 S3.  Our observations indicate there is evidence for two $^{12}$CO\,$(3-2)$ outflows located near IRAS~27, supporting the recent work of \cite{rei04}.  Figure~\ref{fig4} shows the high velocity outflow emission that was used to help determine outflow orientations.  The figure indicates HV gas traces an outflow oriented southeast (SER) to northwest (NWB).  The northeastern (NEB) and southwestern (SWR) lobes vanish in Figure~\ref{fig4} indicating that they have similar velocities and describe a second outflow.  

Figure~\ref{fig5} is a model sketch of the IRAS~27 system, showing the orientations of the $^{12}$CO\,$(3-2)$ outflows.  H$_{2}$, H$\alpha$, and [SII] images of the region reveal an IR source surrounded by a reflection nebula (thick, gray contour) and two giant HH outflows, HH~576 and 577 emanating from a core containing IRAS~27 \citep{rei04}.  Figures~\ref{fig3} and \ref{fig5} show submillimeter cores located along a dense ridge of $870 \, \micron$ emission with molecular outflows oriented at large angles (nearly perpendicular) to the continuum ridge.  It is difficult to determine the source of the $^{12}$CO\,$(3-2)$ outflows, given their proximity to the continuum cores and IRAS object.  It is possible that both outflows originate from IRAS~27.  

The axis of the NEB$-$SWR flow is not as well centered on IRAS~27 as the SER$-$NWB outflow.  An order of magnitude calculation, similar to the analysis in Section (3.2.1), reveals the volume density of the environment, $n_{H_{2}}$, is less than the density required to redirect the outflow ($n_{cl} \sim \, 10^{6}$~cm$^{-3}$).  The alignment of HH and $^{12}$CO\,$(3-2)$ outflows prompted suggestions that IRAS~27 is a binary source with each component driving an outflow \citep{rei04}.  If so, then following the same procedure as for IRAS~25, we estimate $\sim \, 1$~M$_{\odot}$ for the mass of the stellar object(s) associated with IRAS~27.                  

\subsubsection{Outflow Analysis}
  
\subsubsubsection{Optical Depth, Excitation Temperature, \& H$_{2}$ Column Density}

Outflow characteristics and energetics analyses were carried out under assumptions of constant excitation throughout the mapped regions and Local Thermodynamic Equilibrium (LTE).  With pointed observations toward IRAS~25 \& 27 in $^{12}$CO\,$(3-2)$, $^{13}$CO\,$(3-2)$, and $^{12}$CO\,$(1-0)$, measurements of optical depth and excitation temperature were made and H$_{2}$ column density was derived.  Optical depth, $\tau$, was found using $^{13}$CO\,$(3-2)$ and $^{12}$CO\,$(3-2)$ observations.  Assuming $^{12}$CO $\&~^{13}$CO have comparable excitation temperatures due to their similarity in rotational energy level structures, the ratio of their optical depths is the abundance ratio, $r_{a}$ (see Equation~\ref{eq:R1used}).  
\newline
\begin{equation}
\label{eq:R1used}
\frac{I_{^{12}CO}}{I_{^{13}CO}}=\frac{1-e^{-r_{a}\tau_{32}^{^{13}CO}}}{1-e^{-\tau_{32}^{^{13}CO}}}
\end{equation}
\noindent
Optical depth of the $^{13}$CO was determined by making a ratio of the integrated intensities of the $^{12}$CO $\&~^{13}$CO spectra, $I_{^{12}CO}$ and $I_{^{13}CO}$, respectively.  An abundance ratio $r_{a}=60$ for $^{12}$CO to $^{13}$CO was used \citep{kulesa03}, Equation~\ref{eq:R1used} was solved numerically.

After convolving $^{12}$CO\,$(3-2)$ and $^{12}$CO\,$(1-0)$ data sets to the same beamsize ($\sim\,$1$'$) and converting each temperature scale into $T_{mb}$ as described in Section (2.2), line ratios of the spectra were used to determine $T_{ex}$ \citep{walker91}.  In this analysis, the $^{12}$CO\,$(3-2)$ and $^{12}$CO\,$(1-0)$ emission was assumed to originate from the same volume and since our data sets were convolved to the same beamsize, the emission filling factors were the same.  With this method, values of 20~K and 30~K were found for $T_{ex}$ toward IRAS~25 and 27, respectively.

With estimates of optical depth and excitation temperature, the $^{13}$CO\,$(3-2)$ column density was determined with the following expression \citep{groppi04, walker88}:
\newline
\begin{equation}
\label{eq:Nvthin}
N_{\nu,\,thin}=T_{ex} \tau \Delta v ~ \frac{6k}{8 \pi^{3} \nu \mu^{2}}~ \frac{2l+1}{2u+1} ~ \frac{e^{\,lhv/(2kT_{ex})}}{(1-e^{\,-h\nu/(kT_{ex})})}
\end{equation}
\noindent
In the above equation, $l$ is the lower rotational state, $u$ describes the upper state, $k$ and $h$ are respectively Boltzmann's and Planck's constants, $\nu$ is the $^{13}$CO\,$(3-2)$ frequency, $\Delta v$ is the velocity interval over which the column density was calculated, and $\mu$ is the molecular dipole moment expressed in esu.  The above expression describes how to calculate the column density toward a position where the values of $\tau~\&~T_{ex}$ are known.  Pointed $^{13}$CO\,$(3-2)$ observations were made toward individual IRAS sources, though $^{13}$CO\,$(3-2)$ data were not available over the regions mapped in $^{12}$CO\,$(3-2)$.  For this reason, excitation temperature, optical depth, and column density values were obtained at the positions of the IRAS sources, although calculating this information directly over the mapped regions was not possible.  Table~\ref{tab:lteofprops} presents LTE properties of IRAS~25 and 27 calculated for red and blue line wings.    

In order to obtain column density information and ultimately mass and energetics information for outflows NGC~2264~O \& H an emission-weighted area technique was used \citep{lada85}.  The analysis assumed the observed intensity ratio of $^{12}$CO\,$(3-2)$ and $^{13}$CO\,$(3-2)$ emission at the positions of the IRAS sources was constant over the regions mapped in $^{12}$CO\,$(3-2)$.  It was also assumed that $T_{ex}$ obtained at the positions of IRAS sources was constant over the mapped areas.  The intensities of the $^{12}$CO\,$(3-2)$ emission measured toward the IRAS sources, where $N_{\nu,\,thin}$ was calculated, were compared with $^{12}$CO\,$(3-2)$ emission over the entire maps and used to weight the contribution of individual map positions to the total emission weighted area of the outflow, $A_{weight}$.  This procedure is described in Equation~\ref{eq:Aweight}, where $A_{beam}$ is the area of the telescope beam at 345\,GHz, $I_{x,y}$ is the $^{12}$CO\,$(3-2)$ intensity at any map position, and $I_{IRAS~25/27}$ is the $^{12}$CO\,$(3-2)$ intensity at IRAS~25 or 27, depending on the map examined.    
\newline
\begin{equation}
\label{eq:Aweight}
A_{weight}=\sum_{x,y}^{map} A_{beam} (\frac{I_{x,y}}{I_{IRAS~25/27}})
\end{equation}
\noindent
\newline
\begin{equation}
\label{eq:massOF}
M_{OF}=\frac{N_{\nu~thin}}{r_{a}r_{H_{2}}} ~ A_{weight} ~ m_{H_{2}} ~ f_{thin}
\end{equation}

Using the emission-weighted area method, the gas mass associated with the outflow was obtained using Equation~\ref{eq:massOF}.  In this equation, $m_{H_{2}}$ is the molecular hydrogen mass and $f_{thin}$ is the $^{13}$CO\,$(3-2)$ gas filling factor.  The filling factor was calculated using $T_{ex}$, the integrated intensity and optical depth of the $^{13}$CO\,$(3-2)$, and assuming LTE conditions.  Filling factors obtained toward IRAS sources appear in Table~\ref{tab:lteofprops}.  Since $^{13}$CO\,$(3-2)$ measurements were made toward single positions, $f_{thin}$ was assumed to be constant over the maps.  In Equation~\ref{eq:massOF}, $r_{a}$ is the $^{12}$CO$/^{13}$CO abundance ratio.  We adopted $r_{a}=60$ in agreement with the work of \cite{kulesa03}.  The CO$/$H$_{2}$ abundance, $r_{H_{2}}=1\times10^{-4}$, is representative of the ISM and many nearby molecular clouds.  Results of our mass analysis for the red and blue NGC~2264~O \& H outflow lobes are provided in Table~\ref{tab:OFenergetics}. 
  
\subsubsubsection{Outflow Energetics}

With outflow masses, we were able to derive other physical properties of NGC~2264~O \& H, including momentum ($P_{OF}=M_{OF}~ v_{max}$), kinetic energy ($E_{OF}= 1/2~ M_{OF} ~ v_{max}^{2}$), mechanical luminosity ($L_{OF}= E_{OF} / t_{dyn}$), mass outflow rate ($\dot{M}_{OF}= M_{OF} / t_{dyn}$), and force ($\dot{P}_{OF}= P_{OF} / t_{dyn}$).  In this analysis, $t_{dyn}=l_{wing}/v_{max}$ is the outflow dynamical age, $l_{wing}$ is the outflow lobe physical extent measured from emission half-power contours, and $v_{max}$ is the characteristic velocity of the outflowing material.  LTE outflow energetics depend largely upon the choice of $v_{max}$ which is subject to uncertainty due to projection effects of the outflow with respect to the line of sight.  Determining a characteristic velocity representative of the bulk of the outflowing gas is therefore difficult.  For this reason, lower and upper limits of the outflow energetics were determined using outer line wings of $^{13}$CO\,$(3-2)$ (LV RW \& LV BW) and $^{12}$CO\,$(3-2)$ (F~RW \& F~BW) for $v_{max}$.  The results of these calculations appear in Table~\ref{tab:OFenergetics}. 
        
These energetics results are largely consistent with those of \cite{mls88}.  Upper limit kinetic energy, momenta, and mechanical luminosities of the outflows are comparable to those typically associated with young high mass protostars, possessing L$_{FIR}=10^{2}-10^{5}$~L$_{\odot}$ and $M_{OF}=$ tens of M$_{\odot}$ \citep{qzhang05}.  Since neither IRAS~25 nor 27 is believed to be a high mass star forming object, these results are consistant with the interpretation that the outflows arise from multiple objects.  

\section{Analysis \& Discussion}

\subsection{Velocity Centroid Analysis}

\begin{figure}
\includegraphics[scale=0.35]{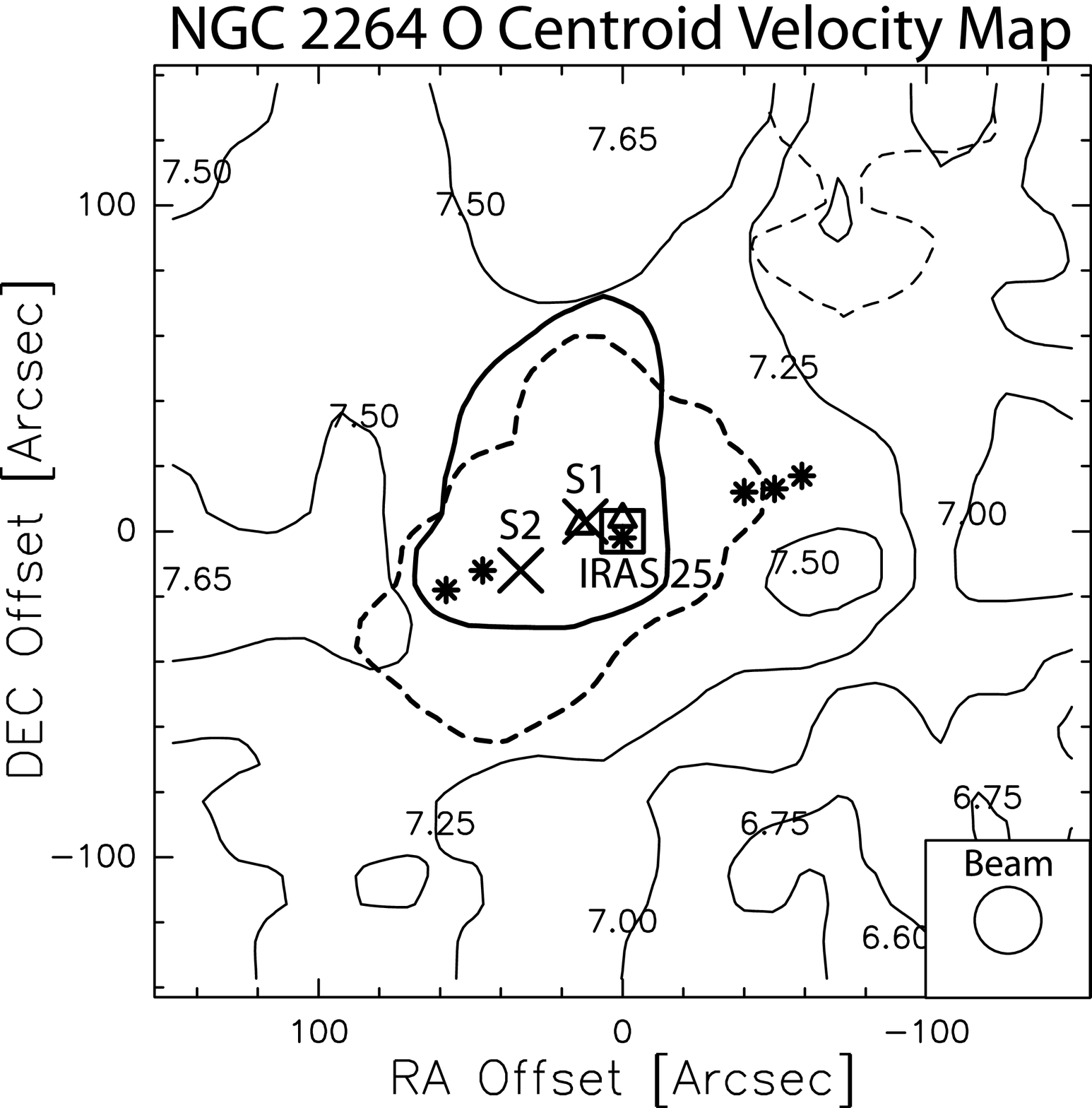}
\includegraphics[scale=0.35]{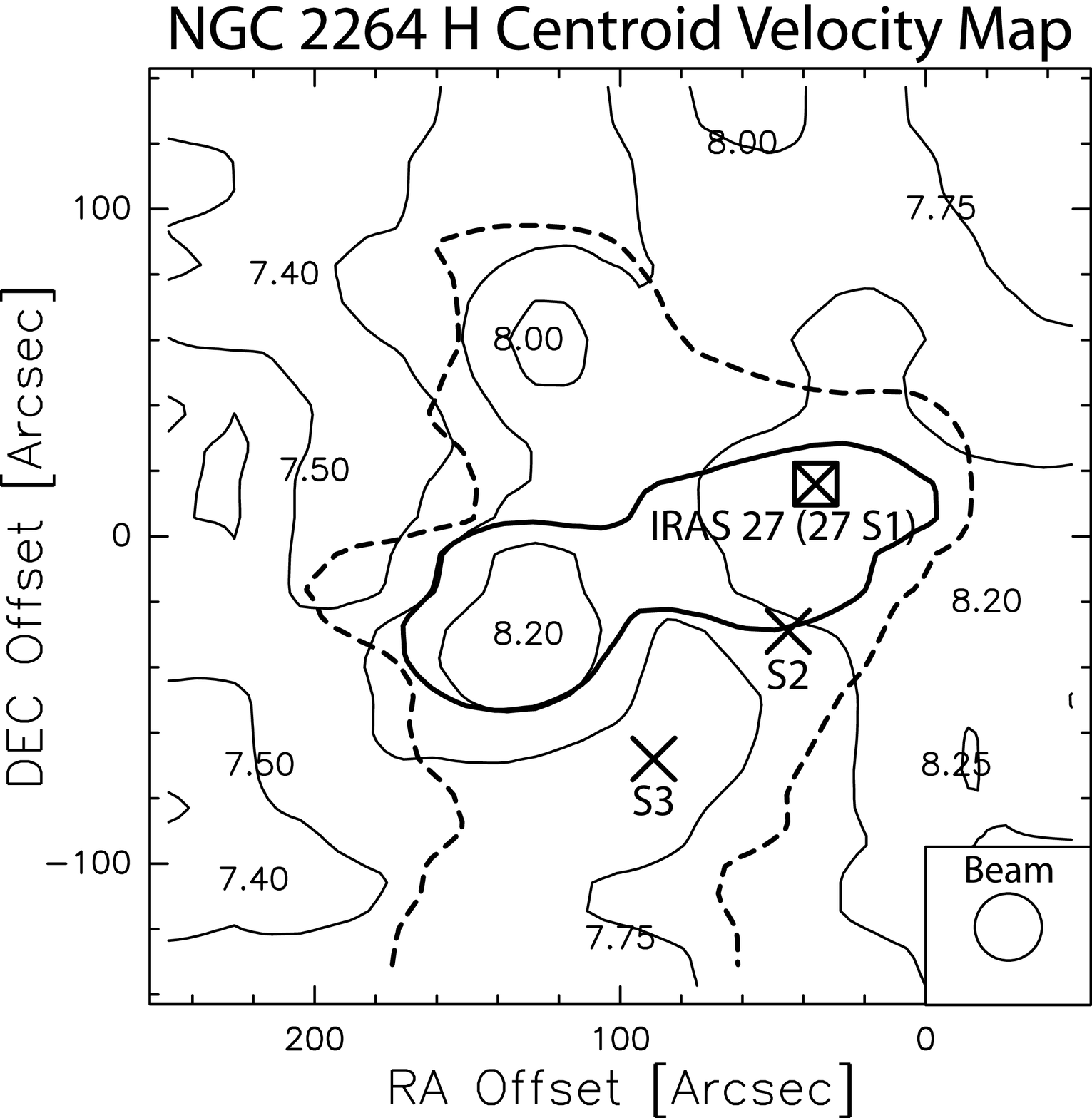}
\caption{\label{fig6} Velocity centroid maps of $^{12}$CO\,$(3-2)$ line core emission associated with NGC~2264~O \& H.  Velocity centroid contours (thin solid lines) are shown and labeled (km s$^{-1}$).  Half power contours of red and blue outflow emission are illustrated with thick solid and dashed contours, respectively.}
\end{figure}

In order to explore the effect the NGC~2264~O \& H molecular outflows have on the cloud environment, we constructed velocity centroid maps of the $^{12}$CO\,$(3-2)$ gas.  The velocity centroid is the value that equally divides the integrated intensity of a line profile.  For this analysis, the velocity centroid was computed over the line core excluding emission in the wings (see Table~\ref{tab:linewings}).  A map of velocity centroid as a function of position was generated for NGC~2264~O \& H (see Figure~\ref{fig6} a and b).  The centroid velocity values have been represented with labeled, thin contours.  Half power contours of the red and blue wing $^{12}$CO\,$(3-2)$ emission are depicted with thick solid and dashed lines, respectively.  There is a small gradient in velocity centroid values in each map, $\Delta v_{cent}=(7.65-6.6)$ km s$^{-1}$ $\approx 1.1$ km s$^{-1}$ and $\Delta v_{cent}=(8.35-7.4)$ km s$^{-1}$ $\approx 1$ km s$^{-1}$ for NGC~2264~O \& H, respectively.  There is no noticeable sign of rotation perpendicular to outflow axes to suggest the signature of an accretion disk.  The centroid velocity gradients extend from the northeast corner of the plots to the southwest, roughly following the orientations of the outflow lobes.  This suggests that NGC~2264~O \& H outflows may dominate the energetics of their surroundings.  

In order to illustrate this quantitatively, the force of bulk motions of cloud material was calculated and compared with the force of the molecular outflow, $\dot{P}_{OF}$.  For this analysis, the mass of the cloud was computed over the line core for each $^{12}$CO\,$(3-2)$ map, $M_{c}$, using the emission weighted area approach described in Section (3.2.3).  This mass calculation assumes the $^{13}$CO\,$(3-2)$ emission is optically thin over the line core.  In order to assess the impact this might have on the analysis, the $^{13}$CO\,$(3-2)$ optical depth was examined as a function of velocity toward the position of the IRAS sources, where direct calculations of optical depth and column density were made.  For IRAS~25, the $^{13}$CO\,$(3-2)$ emission is marginally thick in the line core, with optical depth $\sim\,1-1.1$.  The mass associated with line core velocities, and therefore the corresponding force calculations, should serve as a lower limits for NGC~2264~O.  Over the line core region, the $^{13}$CO\,$(3-2)$ emission toward IRAS~27 is thin, with optical depth values $\sim\,0.3-0.35$, so the mass associated with the line core, and therefore the corresponding force calculations, are most likely valid.           

Calculations of the force of bulk cloud motions were made using $\dot{P}_{c}= M_{c}\, \Delta v_{cent} / \tau_{dyn}$, where $M_{c}$ is the mass in the line core region, $\Delta v_{cent}$ is the observed centroid velocity gradient along the outflow axis, and $\tau_{dyn}$ is the dynamical age of the outflow.  Using dynamical ages of the red and blue outflow lobes resulted in $\dot{P}_{c} = 6.6-11 \times 10^{-4}\, $M$_{\odot}\, $yr$^{-1}$ km s$^{-1}$ for NGC~2264~O.  This is more than $\sim$\,5 times smaller than the force of the outflowing material, $\dot{P}_{OF} = 60 \times 10^{-4}\, $M$_{\odot}\, $yr$^{-1}$ km s$^{-1}$, obtained by summing the contributions of the red and blue wings shown in Table \ref{tab:OFenergetics}.  These comparisons were also made for energetics and dynamical ages obtained using the $^{13}$CO\,$(3-2)$ line wings to guide the analysis.  Values of $\dot{P}_{c} = 3.7-6.2 \times 10^{-4}\, $M$_{\odot}\, $yr$^{-1}$ km s$^{-1}$ were obtained for NGC~2264~O and are $\sim$\,3 times smaller than $\dot{P}_{OF} = 18 \times 10^{-4}\, $M$_{\odot}\, $yr$^{-1}$ km s$^{-1}$.  Overall, the values of $\dot{P}_{c}$ and $\dot{P}_{OF}$ are similar within factors of a few, indicating that the NGC~2264~O outflow has enough force to drive the bulk motions of the cloud material.    

Performing a similar analysis for NGC~2264~H results in $\dot{P}_{c} = 3.7-7.3 \times 10^{-4}\, $M$_{\odot}\, $yr$^{-1}$ km s$^{-1}$.  Compared with the force of the outflow, $\dot{P}_{OF} = 18 \times 10^{-4}\, $M$_{\odot}\, $yr$^{-1}$ km s$^{-1}$, $\dot{P}_{c}$ is several times smaller.  Calculations made using line wings of the $^{13}$CO\,$(3-2)$ emission for the outflow ages and energetics, find values of $\dot{P}_{c} = 2.1-2.2 \times 10^{-4}\, $M$_{\odot}\, $yr$^{-1}$ km s$^{-1}$, similar to the force of the outflow, $\dot{P}_{OF} = 2.4 \times 10^{-4}\, $M$_{\odot}\, $yr$^{-1}$ km s$^{-1}$.  Values of $\dot{P}_{c}$ and $\dot{P}_{OF}$ are within an order of magnitude, similar to NGC~2264~O.  Again it appears that the outflow force is capable of driving bulk motions in the cloud.  
  
\subsection{Energetics Comparison}

\subsubsection{Turbulent Energy}

The turbulent energy contributed to the cloud by the NGC~2264~O \& H outflows, $E_{turb}$, was estimated using our $^{12}$CO\,$(3-2)$ data (observations of an optically thin species, such as C$^{18}$O, were not available).  The estimate was made by comparing average FWHM velocity values at outflow wing emission peaks, $\Delta v($on\,OF$)$, to ambient cloud regions away from the outflow, $\Delta v($Ambient$)$.  In all regions, the line wings were excluded from the Gaussian fit.  Since the $^{12}$CO\,$(3-2)$ is optically thick, opacity broadening due to saturation effects should be considered \citep{phillips79}.  Compared to an optically thin species, saturation is expected to broaden the FWHM of the $^{12}$CO lines by factors of 2.6 and 2.2 toward IRAS~25 and 27, respectively.  Normalizing our measured FWHM values by these factors and assuming that the $^{12}$CO opacity is to first order constant over the mapped regions, we derive values of $\Delta v($on\,OF$)=0.81$ km\,s$^{-1}$ and $\Delta v($Ambient$)=0.54$ km\,s$^{-1}$ for $^{12}$CO\,$(3-2)$ emission surrounding IRAS~25.  Normalized FWHM values of $\Delta v($on\,OF$)=0.86$ km\,s$^{-1}$ and $\Delta v($Ambient$)=0.45$ km\,s$^{-1}$ were obtained for IRAS~27.   

FWHM velocity values measured toward peak outflow emission probe many cloud characteristics including turbulent velocity structures ($v_{turb}$), thermal broadening ($kT/m$), and potentially components due to cloud magnetic field ($v_{B}$):  $\Delta v($on\,OF$) = (v^{2}_{turb}\, +\, kT/m \,+\, v^{2}_{B})^{1/2}$.  The turbulence associated with the FWHM velocity of the outflowing gas, $v_{turb}$, has two components.  These are general cloud turbulence, $v_{turb~cloud}$, present both toward and away from the outflow, and turbulence due to the outflow itself, $v_{turb~OF}$.  These components combine quadratically, $v^{2}_{turb} = v^{2}_{turb~cloud} + v^{2}_{turb~OF}$.  The FWHM velocity measured toward ambient cloud regions away from outflow emission is $\Delta v($Ambient$) = (v^{2}_{turb~cloud}\, +\, v^{2}_{B} \,+\, kT/m)^{1/2}$.  Combining these equations and correcting for opacity broadening effects, the turbulent velocity due to outflow motions, $v_{turb~OF}$, was obtained from FWHM velocity measurements:  $v^{2}_{turb~OF} = [\Delta v($on\,OF$)]^{2} - [\Delta v($Ambient$)]^{2}$.  The turbulent energy contributed to the cloud by the molecular outflow was found using $E_{turb}= 1/2~ M_{OF}~ v_{turb~OF}^{2}$ where $M_{OF}$ is the total outflow mass, obtained by summing contributions of red and blue outflow lobes (see Table~\ref{tab:OFenergetics}).  For regions surrounding IRAS~25 and 27, we found $E_{turb} = 0.75\,$M$_{\odot}\, $km$^{2}\, $s$^{-2}$ and 0.37\,M$_{\odot}\, $km$^{2}\, $s$^{-2}$, respectively.   

The coupling efficiency of outflow energy to the ambient cloud was estimated by comparing the turbulent cloud energy contributed by outflows, $E_{turb}$, and outflow kinetic energy, $E_{OF}$ (see Table~\ref{tab:cloudcoreNRG}).  Values of $E_{OF}$ are orders of magnitude larger than $E_{turb}$, resulting in coupling efficiencies of $\le\,0.5\%$ in the regions surrounding IRAS~25 and 27.  The low coupling efficiency of molecular outflow energy to the surrounding environment together with the results of our centroid velocity analysis suggests that most of the outflows' energy is directed along the flow axis and deposited in the intercloud medium leaving adjacent regions of the parent cloud largely undisturbed.    

\subsubsection{Gravitational Binding Energy}

Clump gravitational binding energy, $E_{GC}$, was obtained using results of the continuum analysis (Table~\ref{tab:coreproperties}) and the expression for gravitational potential energy of a uniform gravitating sphere of constant density, $E_{GC}=3/5~ (G ~ M^{2}_{H_{2}}/R)$.  Here $G$ is the gravitational constant, $M_{H_{2}}$ is the molecular mass, and $R$ is the sphere radius used to compute $M_{H_{2}}$.  For IRAS~25, the average clump radius is $\sim\,34''$ and substitution yields $E_{GC} = 640 $M$_{\odot}\, $km$^{2}\, $s$^{-2}$.  

The continuum image of the region surrounding IRAS~27 contains three sources, 27~S1, 27~S2, and 27~S3.  Core masses, $M_{H_{2}}$, derived from $870\,\micron$ flux measurements were used to individually determine contributions to $E_{GC}$ for the three sources.  Since the core masses, $M_{H_{2}}$, were derived from fluxes observed within $45''$ apertures, $R = 22.5''$ was used to obtain $E_{GC}$ of individual cores.  The results were summed to estimate the total clump gravitational potential energy:  $E_{GC} = 5.5 $M$_{\odot}\, $km$^{2}\, $s$^{-2}$.  The result is included in Table~\ref{tab:cloudcoreNRG} and is a lower limit to $PE_{GC}$ since individual sources lie along a ridge of extended low-intensity $870\,\micron$ emission.  The mass associated with the low-level emission was not found since the FIR SED analysis did not apply. 

Table~\ref{tab:cloudcoreNRG} compares outflow kinetic energy, turbulent energy contributed to the cloud by outflows, and cloud clump potential energy for the regions surrounding IRAS~25 and 27.  The molecular cloud appears to remain gravitationally bound, since turbulent energy injected into the cloud by outflows, $E_{turb}$, is much smaller than the cloud clump potential energy, $E_{GC}$.  The low outflow energy coupling efficiency may explain why, even in the presence of multiple outflows, the cloud cores appear to maintain their overall integrity, with $E_{GC} \gg E_{turb}$.  The dismantling effects the NGC~2264~O \& H outflows are having on the molecular cloud appear to be largely localized to the portions of the clouds contained within opening angles of the outflow lobes.    

\section{Summary}

We have made $870 \, \micron$ continuum measurements and $^{12}$CO\,$(3-2)$, $^{13}$CO\,$(3-2)$, and $^{12}$CO\,$(1-0)$ observations of IRAS~25 \& 27 in the northern cloud complex of NGC~2264 using the HHT and 12m telescopes.  Continuum $870 \, \micron$ OTF maps (5$'$$\times$5$'$) were made toward the sources IRAS~25 (IRAS 06382+1017) \& 27 (IRAS 06381+1039) and several continuum cores were identified.  SEDs were constructed for the cores and used to derive column densities, gas masses, FIR luminosities, and dust temperatures.  A molecular line and energetics analysis of the outflows associated with NGC~2264~O \& H was performed and multiple outflows were found within each region.  In order to assess the impact star formation is having on the NGC~2264 cloud, the turbulent energy, outflow dynamical energy, and cloud clump gravitational potential energy were compared.  The results are consistent with the interpretation that the cloud complexes are maintaining their overall integrity except along outflow axes where cloud material directly interacts with the outflows.  The outflows deposit most of their energy outside of the molecular cloud, resulting in a weak $\le\,0.5\%$ coupling between outflow kinetic energy and cloud turbulent energy.

\acknowledgments

We are grateful to C. Kulesa, G. Wolf-Chase, B. Vila-Vilaro, and the Arizona Radio Observatory staff and telescope operators for their helpful discussions and aid in assembling this data set.  This work was supported by NASA GSRP grant number 6267.
  

\clearpage

\begin{deluxetable}{ccccccccccccc}
\tabletypesize{\scriptsize}
\rotate
\tablecaption{Spectral Line Observations \label{tab:observations}}
\tablewidth{0pc}
\tablehead{
\colhead{Object} & \colhead{$\alpha_{1950}$} & \colhead{$\delta_{1950}$} &
\colhead{Telescope} & \colhead{Spectral Line} & \colhead{Dates} & 
\colhead{Mode} & \colhead{Map Size} & \multicolumn{2}{c}{Resolution}
& \colhead{$2\sigma$~RMS} & \multicolumn{2}{c}{Beam Size} \\ 
\colhead{} & \colhead{} & \colhead{} & \colhead{} & \colhead{} & 
\colhead{} & \colhead{} & \colhead{($\,''$)} &
\colhead{(km s$^{-1}$)} & \colhead{(kHz)} & \colhead{(K)} & 
\colhead{($\,''$)} & \colhead{(pc)} }
\startdata
IRAS 25 & 06:38:17.0 & 10:18:00.0 & HHT & $^{12}$CO($3-2$) &
2004 Feb 25-27 & OTF & 300$\times$300 &
0.886 & 1021 & 0.3 & 22 & 0.08 \\
(NGC 2264 O) & & & & & & & & & & & & \\
IRAS 25 & 06:38:17.0 & 10:18:00.0 & HHT & $^{13}$CO($3-2$) &
2004 Feb 25-27 & APS & \nodata &
0.927 & 1021 & 0.06 & 23 & 0.09 \\
(NGC 2264 O) & & & & & & & & & & & & \\
IRAS 25 & 06:38:17.0 & 10:18:00.0 & 12m & $^{12}$CO($1-0$) &
2002 Apr 08-10 & APS & \nodata & 0.254 & 97.6 
& 0.1 & 55 & 0.21 \\
(NGC 2264 O) & & & & & & & & & & & & \\
NGC 2264 H & 06:38:16.8 & 10:39:45.0 & HHT & $^{12}$CO($3-2$) &
2004 Feb 25-27 & OTF & 300$\times$300 &
0.886 & 1021 & 0.22 & 22 & 0.08 \\
  & & & & & & & & & & & & \\
IRAS 27 & 06:38:13.0 & 10:39:45.0 & HHT & $^{13}$CO($3-2$) &
2004 Feb 25-27 & APS & \nodata &
0.927 & 1021 & 0.06 & 23 & 0.09 \\
  & & & & & & & & & & & & \\
IRAS 27 & 06:38:13.0 & 10:39:45.0 & 12m & $^{12}$CO($1-0$) &
2002 Apr 08-10 & APS & \nodata & 0.254 & 97.6 & 1.2 & 55 & 0.21 \\
\enddata
\end{deluxetable}

\clearpage

\begin{deluxetable}{ccccc}
\tabletypesize{\scriptsize}
\tablecaption{870 \micron~Source Flux \label{tab:870srcflux}}
\tablewidth{0pc}
\tablehead{
\colhead{Source} & \colhead{$\alpha_{1950}$} & \colhead{$\delta_{1950}$} &
\colhead{$45''$ Aperture Flux} & \colhead{$3'\times5'$ Aperture Flux} \\ 
\colhead{} & \colhead{} & \colhead{} & \colhead{(Jy)} & 
\colhead{(Jy)} }
\startdata
IRAS 25 & 06:38:17.0 & 10:18:00.0 & 2.60$\pm$0.52 & 8.29$\pm$1.66 \\
25 S1 & 06:38:17.8 & 10:18:02.6 & 6.36$\pm$1.27 & \nodata \\
25 S2 & 06:38:19.2 & 10:17:47.7 & 5.99$\pm$1.20 & \nodata \\
IRAS 27 (27 S1) & 06:38:13.0 & 10:39:45.0 & 3.81$\pm$0.64 & 22.38$\pm$4.48 \\
27 S2 & 06:38:13.0 & 10:39:01.0 & 2.24$\pm$0.45 & \nodata \\
27 S3 & 06:38:15.9 & 10:38:22.5 & 1.96$\pm$0.39 & \nodata \\
\enddata
\end{deluxetable}


\begin{deluxetable}{cccccc}
\tabletypesize{\scriptsize}
\tablecaption{Source Flux \label{tab:srcflux}}
\tablewidth{0pc}
\tablehead{
\colhead{Source} & \colhead{Wavelength} & \colhead{Flux} &
\colhead{$\sigma$~RMS} & \colhead{Aperture} & \colhead{References} \\ 
\colhead{} & \colhead{(\micron)} & \colhead{(Jy)} & \colhead{(Jy)} & 
\colhead{($\,''$)} & \colhead{} }
\startdata
IRAS 25 & 870 & 8.29 & 1.66 & 180$\times$300 & 1 \\
        & 850 & 6.25 & 1.25 & 180$\times$300 & 2 \\
        & 450 & 21.9 & 4.38 & 180$\times$300 & 2 \\
        & 100 & 86.66 & 30.33 & 180$\times$300 & 3 \\
        & 60 & 43.45 & 8.69 & 90$\times$285 & 3 \\
        & 25 & 7.908 & 0.949 & 45$\times$276 & 3 \\
        & 12 & 1.283 & 0.192 & 45$\times$270 & 3 \\
IRAS 27 (27 S1) & 1300 & 0.8 & 0.16 & 27.3 & 2 \\
        & 870 & 3.81 & 0.76 & 45.0 & 1 \\
        & 850 & 3.20 & 0.64 & 45.0 & 2 \\
        & 450 & 31.8 & 6.36 & 45.0 & 2 \\
        & 100 & 144.9 & 37.67 & 180$\times$300 & 3 \\
        & 60 & 53.66 & 11.81 & 90$\times$285 & 3 \\
        & 25 & 2.755 & 0.193 & 45$\times$276 & 3 \\
        & 12 & 0.25 & 0.001 & 45$\times$270 & 3 \\
27 S2  & 1300 & 0.7 & 0.14 & 27.3 & 2 \\
        & 870 & 2.24 & 0.45 & 45.0 & 1 \\
        & 850 & 1.87 & 0.37 & 45.0 & 2 \\
        & 450 & 12.1 & 2.42 & 45.0 & 2 \\
        & 100 & 144.9 & 37.67 & 180$\times$300 & 3 \\
        & 60 & 53.66 & 11.81 & 90$\times$285 & 3 \\
        & 25 & 2.755 & 0.193 & 45$\times$276 & 3 \\
        & 12 & 0.25 & 0.001 & 45$\times$270 & 3 \\
27 S3   & 870 & 1.96 & 0.39 & 45.0 & 1 \\
        & 850 & 1.72 & 0.34 & 45.0 & 2 \\
        & 450 & 12.9 & 2.58 & 45.0 & 2 \\
        & 100 & 144.9 & 37.67 & 180$\times$300 & 3 \\
        & 60 & 53.66 & 11.81 & 90$\times$285 & 3 \\
        & 25 & 2.755 & 0.193 & 45$\times$276 & 3 \\
        & 12 & 0.25 & 0.001 & 45$\times$270 & 3 \\
\enddata
\tablecomments{References:  (1) This work.  (2) \cite{wc03}  (3) IRAS PSC}
\end{deluxetable}

\clearpage

\begin{deluxetable}{cccccccc}
\tabletypesize{\scriptsize}
\tablecaption{Parameters of FIR SED Model Fits \label{tab:sedparms}}
\tablewidth{0pc}
\tablehead{
\colhead{Source} & \colhead{T} & \colhead{$\tau_{870}$} &
\colhead{$\beta$} & \colhead{$\Omega_{m}$} & \colhead{$\Omega_{s}$} &
\colhead{$f_{d}$} & \colhead{$L_{FIR}$} \\ 
\colhead{} & \colhead{(K)} & \colhead{} & \colhead{} & \colhead{} & 
\colhead{} & \colhead{} & \colhead{(L$_{\odot}$)} }
\startdata
IRAS 25 & 10 & $1.3 \times 10^{-2}$ & 1.4 & $1.2 \times 10^{-8}$ & 
$2.4 \times 10^{-9}$ & $21 \times 10^{-2}$ & 88 \\
        & 40 & $1.5 \times 10^{-3}$ & 1.4 & $1.2 \times 10^{-8}$ & 
$2.0 \times 10^{-10}$ & $1.7 \times 10^{-2}$ & 88 \\
IRAS 27 (27 S1) & 30 & $2.7 \times 10^{-3}$ & 1.4 & $1.7 \times 10^{-8}$ & 
$9.3 \times 10^{-10}$ & $5.3 \times 10^{-2}$ & 120 \\
27 S2 & 36 & $1.3 \times 10^{-3}$ & 1.2 & $1.7 \times 10^{-8}$ & 
$3.2 \times 10^{-10}$ & $1.9 \times 10^{-2}$ & 92 \\
27 S3 & 33 & $1.2 \times 10^{-3}$ & 1.4 & $1.7 \times 10^{-8}$ & 
$5.1 \times 10^{-10}$ & $2.9 \times 10^{-2}$ & 100 \\
\enddata
\end{deluxetable}


\begin{deluxetable}{ccccc}
\tabletypesize{\scriptsize}
\tablecaption{Core Properties \label{tab:coreproperties}}
\tablewidth{0pc}
\tablehead{
\colhead{Source} & \colhead{$M_{H_{2}}$} & \colhead{$N_{H_{2}}$} &
\colhead{$n_{H_{2}}$} & \colhead{$A_{v}$} \\ 
\colhead{} & \colhead{(M$_{\odot}$)} & \colhead{(cm$^{-2}$)} & \colhead{(cm$^{-3}$)} & \colhead{} }
\startdata
IRAS 25 & 140 & $3.0 \times 10^{23}$ & $3.0 \times 10^{5}$ & 330 \\
IRAS 27 (27 S1) & 12 & $7.0 \times 10^{22}$ & $8.8 \times 10^{4}$ & 77 \\
27 S2 & 3.4 & $1.9 \times 10^{22}$ & $2.5 \times 10^{4}$ & 21 \\
27 S3 & 5.5 & $3.2 \times 10^{22}$ & $4.0 \times 10^{4}$ & 35 \\
\enddata
\end{deluxetable}


\begin{deluxetable}{cccccccc}
\tabletypesize{\scriptsize}
\tablecaption{CO($3-2$) Line Wings \label{tab:linewings}}
\tablewidth{0pc}
\tablehead{
\colhead{Source} & \colhead{F~BW} & \colhead{HV BW} &
\colhead{LV BW} & \colhead{Line Core} & \colhead{LV RW} & \colhead{HV RW} &
\colhead{F~RW} \\ 
\colhead{} & \colhead{(km s$^{-1}$)} & \colhead{(km s$^{-1}$)} & \colhead{(km s$^{-1}$)} & \colhead{(km s$^{-1}$)} & \colhead{(km s$^{-1}$)} & \colhead{(km s$^{-1}$)} & \colhead{(km s$^{-1}$)} }
\startdata
IRAS 25 & $-10.5,\, 5.6$ & $-10.5,\, -2.5$ & $-2.5,\, 5.6$ & $5.6,\, 8.8$ & 
$8.8,\, 19.0$ & $19.0,\, 27.5$ & $8.8,\, 27.5$ \\
IRAS 27 & $-20.0,\, 6.7$ & $-20.0,\, 0.0$ & $0.0,\, 6.7$ & $6.7,\, 9.8$ & 
$9.8,\, 20.0$ & $20.0,\, 28.5$ & $9.8,\, 28.5$ \\

\enddata
\end{deluxetable}


\begin{deluxetable}{ccccccc}
\tabletypesize{\scriptsize}
\tablecaption{LTE Outflow Analysis Properties \label{tab:lteofprops}}
\tablewidth{0pc}
\tablehead{
\colhead{Source} & \colhead{Line Wing} & \colhead{T$_{ex}$} & \colhead{$\tau_{32}^{^{13}CO}$} & \colhead{$N_{\nu,~thin}$} & \colhead{$N_{H_{2}}$} & \colhead{$f_{thin}$} \\ 
\colhead{} & \colhead{} & \colhead{(K)} & \colhead{} & \colhead{($\times 10^{15}$ cm$^{-2}$)} & \colhead{($\times 10^{21}$ cm$^{-2}$)} & \colhead{} }
\startdata
IRAS~25 & F~RW & 20 & 0.16 & 14 & 8.6 & 0.17 \\
IRAS~25 & F~BW & 20 & 0.28 & 22 & 13 & 0.10 \\
IRAS~27 & F~RW & 30 & 0.10 & 14 & 8.5 & 0.04 \\
IRAS~27 & F~BW & 30 & 0.03 & 6.6 & 3.9 & 0.06 \\
\enddata
\end{deluxetable}

\clearpage

\begin{deluxetable}{lcccccccc}
\rotate
\tabletypesize{\scriptsize}
\tablecaption{NGC~2264 $^{12}$CO($3-2$) Outflow Energetics \label{tab:OFenergetics}}
\tablewidth{0pc}
\tablehead{
\colhead{Outflow} & \colhead{$v_{max}$} & \colhead{$M_{OF}$} & \colhead{$P_{OF}$} & \colhead{$E_{OF}$} & \colhead{$L_{OF}$} & \colhead{$\dot{M}_{OF}$} & \colhead{$t_{dyn}$} & \colhead{$\dot{P}_{OF}$} \\ 
\colhead{} & \colhead{(km\,s$^{-1}$)} & \colhead{(M$_{\odot}$)} & \colhead{(M$_{\odot}\, $km$\, $s$^{-1}$)} & \colhead{($\times 10^{2}\, $M$_{\odot}\, $km$^{2}\, $s$^{-2}$)} & \colhead{(L$_{\odot}$)} & \colhead{($\times 10^{-5}\, $M$_{\odot}\, $yr$^{-1}$)} & \colhead{($\times\,10^{4}$ yr)} & \colhead{($\times 10^{-4}\, $M$_{\odot}\, $yr$^{-1}$ km s$^{-1}$)} }
\startdata
O (LV~RW) & 11 & 1.5 & 16 & 0.9 & 0.81 & 8.1 & 1.8 & 8.1 \\
O (LV~BW) & 10.5 & 2.9 & 31 & 1.6 & 0.89 & 9.8 & 3.0 & 10 \\
O (F~RW) & 19.5 & 1.5 & 29 & 2.8 & 4.5 & 14 & 1.0 & 28 \\
O (F~BW) & 18.5 & 2.9 & 54 & 5.0 & 4.9 & 17 & 1.7 & 32 \\
H (LV~RW) & 12 & 0.59 & 7.1 & 0.43 & 0.13 & 1.1 & 5.4 & 1.3 \\
H (LV~BW) & 8.0 & 0.79 & 6.3 & 0.25 & 0.07 & 1.4 & 5.7 & 1.1 \\
H (F~RW) & 20.5 & 0.59 & 12 & 1.2 & 0.65 & 1.9 & 3.2 & 3.8 \\
H (F~BW) & 28 & 0.79 & 22 & 3.1 & 3.1 & 4.9 & 1.6 & 14 \\
\enddata
\tablecomments{Nomenclature:  O~$=$~NGC~2264~O and H~$=$~NGC~2264~H}
\end{deluxetable}

\clearpage

\begin{deluxetable}{ccccccc}
\tabletypesize{\scriptsize}
\tablecaption{Cloud Core Energetics \label{tab:cloudcoreNRG}}
\tablewidth{0pc}
\tablehead{
\colhead{Source} & \colhead{$M_{H_{2}}$} & \colhead{$M_{OF}$} &
\colhead{$E_{GC}$} & \colhead{$E_{OF}$} & \colhead{$E_{turb}$} &
\colhead{Comparison} \\ 
\colhead{} & \colhead{(M$_{\odot}$)} & \colhead{(M$_{\odot}$)} & \colhead{(M$_{\odot}\, $km$^{2}\, $s$^{-2}$)} & \colhead{(M$_{\odot}\, $km$^{2}\, $s$^{-2}$)} & \colhead{(M$_{\odot}\, $km$^{2}\, $s$^{-2}$)} & \colhead{} }
\startdata
IRAS 25 & 140 & 4.1 & 640 & 430\,$-$\,780 & 0.75 & $E_{GC}\sim 850\,E_{turb}$ \\

IRAS 27 & 21 & 1.4 & 5.5 & 68\,$-$\,250 & 0.37 & $E_{GC}\sim 15\,E_{turb}$ \\
\enddata
\tablecomments{$E_{OF}$ values were determined using $^{13}$CO\,$(3-2)$ and $^{12}$CO\,$(3-2)$ line wings, respectively.}
\end{deluxetable}

\clearpage

\clearpage

\clearpage

\clearpage

\clearpage

\clearpage

\end{document}